\documentclass[a4paper,11pt]{article}
\pdfoutput=1 % if your are submitting a pdflatex (i.e. if you have
\usepackage{jheppub} % for details on the use of the package, please
\usepackage[T1]{fontenc} % if needed

\usepackage{graphicx}
\usepackage{bm}
\usepackage{braket}
\usepackage{hyperref}
\usepackage[usenames]{xcolor}
\usepackage{amsmath}
\usepackage{graphicx}
\usepackage{latexsym}
\usepackage{amsmath,amssymb}
\usepackage{xcolor}
\usepackage{stackrel}
\usepackage{accents}
\usepackage{empheq}
\usepackage[title]{appendix}

\usepackage{float}
\usepackage{braket}
\usepackage{bbold}
\usepackage{mathtools}

\usepackage{float}
\usepackage{braket}
\usepackage{bbold}
\usepackage{mathtools}
\usepackage{mdframed}
\DeclareMathOperator{\sgn}{sgn}

\DeclareMathOperator{\Tr}{Tr}
\DeclareMathOperator{\tr}{tr}

\newcommand{\vex}[1]{\bm{\mathrm{#1}}}

\newcommand{\D}{\mathcal{D}}

\newcommand{\bpsi}{\bar{\psi}}

\newcommand{\ib}{\vex{i}}
\newcommand{\jb}{\vex{j}}
\newcommand{\kb}{\vex{k}}

\newcommand{\re}{\operatorname{Re}}

\newcommand{\Mod}{\,\mathrm{mod}\,}

\title{
Effective field theory of random quantum circuits
}

\author[a,b]{Yunxiang Liao,}
\author[a]{Victor Galitski}

\affiliation[a]{Joint Quantum Institute, Department of Physics, University of Maryland, College Park, MD 20742, USA}
\affiliation[b]{Condensed Matter Theory Center, Department of Physics, University of Maryland, College Park, MD 20742, USA}

\emailAdd{liao@umd.edu}
\emailAdd{galitski@umd.edu}

\abstract{Quantum circuits have been widely used as a platform to simulate generic quantum many-body systems. In particular, random quantum circuits provide a means to probe universal features of many-body quantum chaos and ergodicity. Some such features have already been experimentally demonstrated in the noisy intermediate-scale quantum (NISQ) devices. On the theory side, properties of random quantum circuits have been studied on a case-by-case basis and for certain specific systems, a hallmark of quantum chaos - universal Wigner-Dyson level statistics - has been derived. This work develops an effective field theory for a large class of random quantum circuits. The theory has the form of a replica sigma model and is similar to the low-energy  approach to diffusion in disordered systems.  The method is used to explicitly derive universal random matrix behavior of a large family of random circuits. In particular, we rederive Wigner-Dyson spectral statistics of the brickwork circuit model by Chan, De~Luca, and Chalker [Phys. Rev. X {\bf 8}, 041019 (2018)] and show within the same calculation that its various permutations  and higher-dimensional generalizations preserve the universal level statistics. Finally, we use the replica sigma model framework to rederive the Weingarten calculus, which is a method to evaluate integrals of polynomials of matrix elements with respect to the Haar measure over compact groups and has many applications in the studies of quantum circuits.
The effective field theory, derived here,  provides both a method to quantitatively characterize quantum dynamics of random Floquet systems (e.g., calculating operator  and entanglement spreading) and also a path to understanding  the general fundamental mechanism behind quantum chaos and thermalization in these systems.}

\begin{document} 
\maketitle
\flushbottom

\section{Introduction}~\label{sec:int}

Recent years have seen a surge of interest in random quantum circuits, which can be used to simulate various properties of interacting many-body quantum systems, including universal ergodic dynamics.
Random quantum circuits~\cite{QC-0,QC-1,QC-2,QC-3,QC-4,QC-5,QC-6,QC-7,QC-8,QC-9,QC-10,QC-11}  consist of qubits (or qudits) evolving under successive applications of  unitary quantum gates drawn randomly from ensembles of unitaries.
They capture several general properties of many-body quantum chaotic systems.
In particular, universal random matrix theory (RMT) statistics of the quasi-energy spectra has been derived for some time-periodic (Floquet) circuits~\cite{Chalker-1,Chalker-2,Chalker-3,Chalker-4, Chalker-5,Chalker-6, Prosen-SFF3,Prosen-SFF2,Prosen-GSFF,Huse2021, chan2021}.
Quantum systems with underlying classical chaotic dynamics are conjectured to exhibit level statistics identical to that of a suitably chosen random matrix ensemble~\cite{BGS}, and the RMT statistics has been used as one of the indicators for quantum chaos.
For decades, numerous efforts have been made to understand theoretically this Bohigas-Giannoni-Schmit (BGS) quantum chaos conjecture~\cite{Prosen-SFF1,Prosen-Fermion,Prosen-Boson,Prosen-KIsing,BerrySFF, Sieber_2001, Muller_2005,SYK-Altland,SSS, PRB,PRR,PRL}.
Random quantum circuits, due to their fine-tuned structure, allow exact analysis of the spectral statistics, which sheds light on the underlying mechanism responsible for the emergence of RMT structure.
In addition to the RMT spectral statistics, random quantum circuits also exhibit other fundamental properties of many-body quantum chaotic systems, such as the decay of correlation functions of local observables~\cite{Prosen-DUQC-corr,Prosen-DUQC-dyn}, ballistic spreading of the local operators~\cite{Nahum-2018,Sondhi,Keyserlingk-2,Huse-2018, HJ-2018,Chalker-1,Bertini}, ballistic growth of the entanglement~\cite{Nahum-2017,Nahum-2018-2,Sondhi,Prosen-Spin-EE,Prosen-DUQC-dyn,Prosen-QDUC-EE-I,Prosen-QDUC-EE-II,Chalker-1,Gopalakrishnan,Keyserlingk-2019,Sondhi,Chalker-1}
and Gaussian distribution of the matrix elements of observables in the energy eigenbasis (as expected from the eigenstate thermalization hypothesis~\cite{srednicki1994chaos,srednicki1999,Deutsch})~\cite{Chalker-ETH,Prosen-ETH}.
Experimentally, random quantum circuits can be simulated in the noisy intermediate-scale quantum (NISQ) devices~\cite{Preskill} built with superconducting qudits~\cite{SC-1,SC-2}, trapped ions~\cite{ion-1,ion-2}, and Rydberg atoms~\cite{Rydberg},  and some of these generic features for quantum chaotic systems have been observed~\cite{google-scrambling}.
Due to their high controllability, random quantum circuits provide a useful tool to explore fundamental principles underlying chaos and  thermalization.

At  a high level, the approaches to study the conventional interacting many-body systems can be broadly categorized as either (i)~The microscopic  approach, where the specific  Hamiltonian of a particular system is studied  in the ab initio or bottom up way and where all microscopic details enter the calculation or (ii)~The low-energy (usually field-theory-based) approach where only the universal low-energy features enter the theory (e.g., gapless collective modes). In the context of the quantum circuits used to simulate many-body systems, it was mostly the former-type of approach that has been used to date. 
In particular, theoretical studies of random quantum circuits relied heavily on  fine-tuned features of specific models. Here we construct a field-theory approach, where universal features of ergodic dynamics of a large family of random quantum circuits can be studied in a unified way.

This field-theory method generalizes a supersymmetric sigma model developed by Zirnbauer~\cite{Zirnbauer_1996, Zirnbauer_1998, Zirnbauer_1999}, Altland~\cite{QKR} and others making use of a generalized Hubbard-Stratonovich transformation -
the color-flavor transformation~\cite{Zirnbauer_1996, Zirnbauer_1998, Zirnbauer_1999, Zirnbauer_2021}.
This model is similar to the sigma model for disordered electron systems~\cite{Wegner-1979,Efetov},
but is formulated for systems modeled by  an evolution operator or a scattering matrix, rather than a Hamiltonian. 
In particular, it has been used to investigate the connection between the appearance of universal RMT statistics and the underlying classical chaotic dynamics (i.e., the BGS quantum chaos conjecture~\cite{BGS}) in generic quantum chaotic maps~\cite{Zirnbauer_1998, Zirnbauer_1999}, localization in the quantum kicked rotors~\cite{QKR,Tian_2010,Tian_2011,Tian_2012}, eigenenergy and eigenfunction statistics of quantum graphs~\cite{graph-1,graph-2,graph-eigenfun,graph-eigenfun-2},
and the plateau transition in quantum Hall systems~\cite{Zirnbauer-QH,Zirnbauer-QH2}.
See Refs.~\cite{Altland-rev,Haake} for a review of this model and a more complete list of references.
For quantum circuits, we find it convenient to reformulate this sigma model in the replica  formalism (in contrast to supersymmetry), and use it to derive an ensemble-averaged effective field theory.

\begin{figure}[t!]
	\centering % \begin{center}/\end{center} takes some additional vertical space
	\includegraphics[width=.75\textwidth]{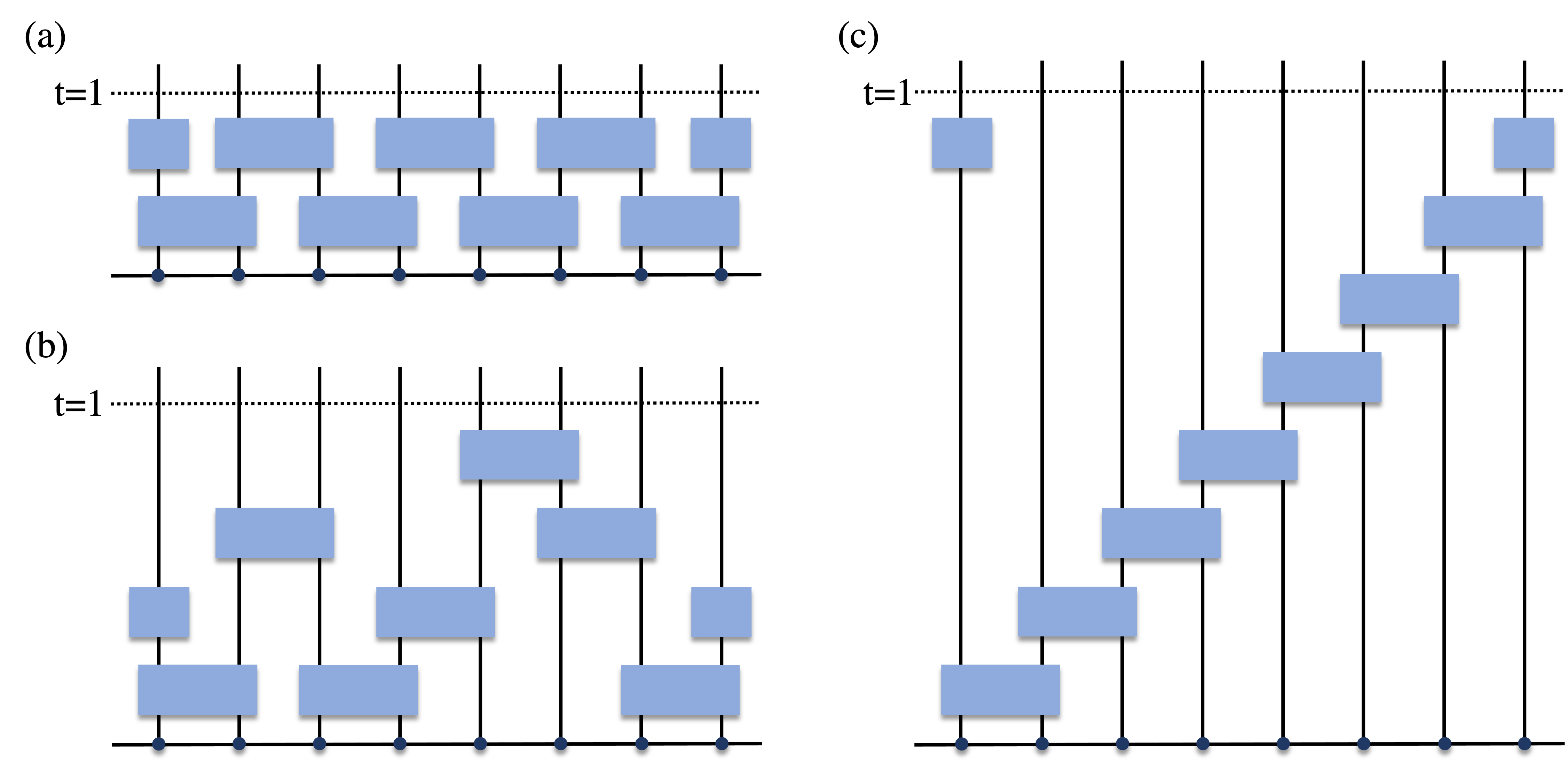}
	\caption{\label{fig:i} Floquet operators of 1D Floquet random quantum circuits with different orderings of the two-qudit unitary gates: (a) brickwork circuit~\cite{Chalker-1}, (c) staircase circuit, and (b) circuit obtained from rearranging the local gates (blue boxes) in (a) or (b). 
	The horizontal direction represents the space coordinates, and each qudit is indicated by a black dot. 
	The vertical direction shows the discrete time evolution within one period, and different layers represent different substeps.
	Each blue box represents an independent CUE matrix $w^{(n,n+1)}$ acting on the Hilbert space of two neighboring sites $n$ and $n+1$. 
	The two half blue boxes at the boundaries constitute a random CUE matrix $w^{(L,1)}$ (identical matrix) acting on the sites $L$ and $1$,  for periodic (open) boundary condition, with $L$ being the total number of qudits. 
	For any of these Floquet quantum circuits, which differ only by the ordering of local gates and are subject to either periodic or open boundary condition, we prove that the statistical properties of the quasi-energy spectra are identical to that of the CUE ensemble in the limit of large onsite Hilbert space dimension $q\rightarrow \infty$.}
\end{figure}

\begin{figure}[t!]
	\centering % \begin{center}/\end{center} takes some additional vertical space
	\includegraphics[width=.85\textwidth]{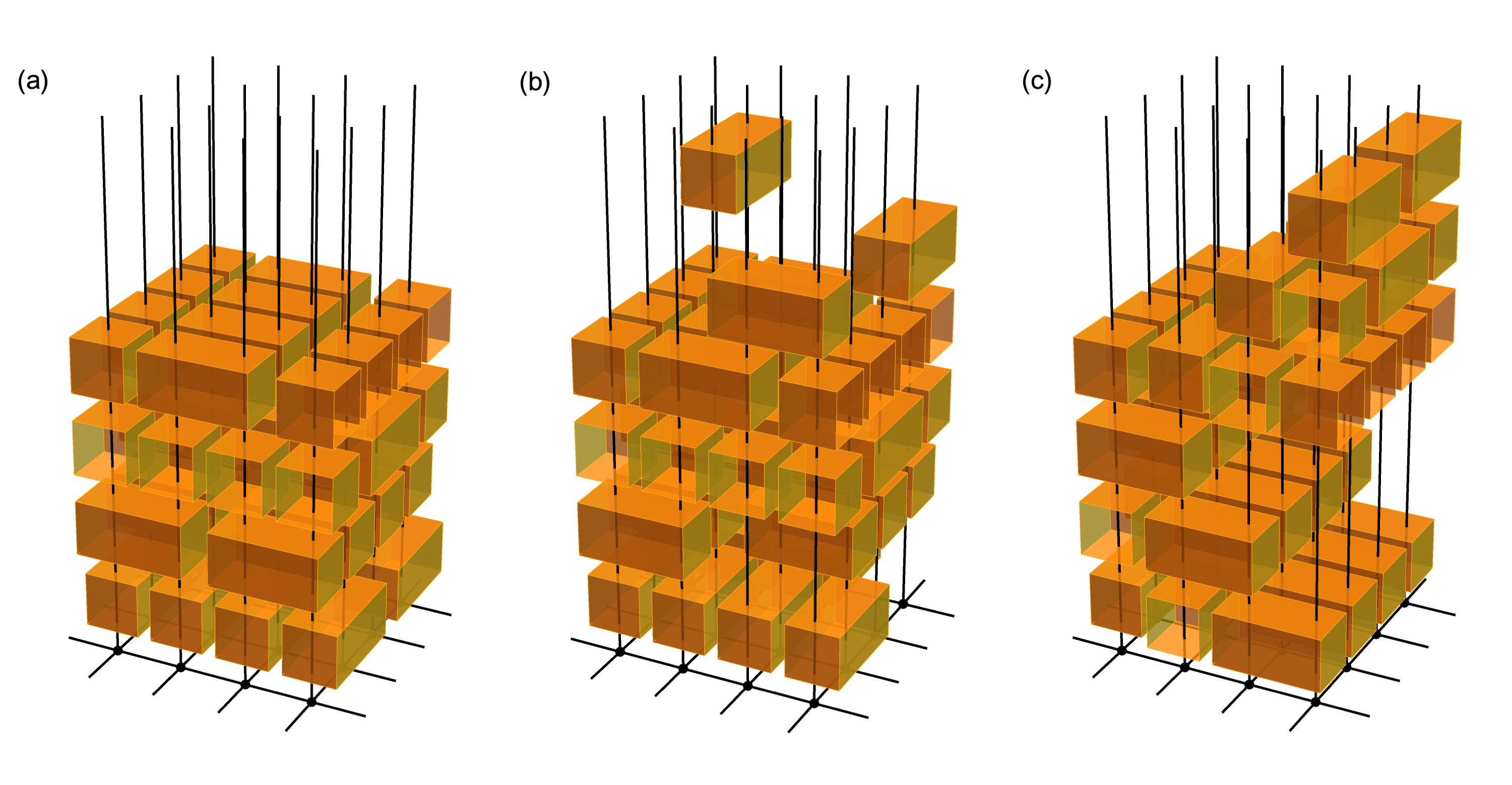}
	\caption{\label{fig:ii} Floquet operators of 2D Floquet random quantum circuits with different orderings of the two-qudit gates.  Within one period, two-qudit unitaries (orange boxes) drawn independently from the CUE ensemble are applied to all pairs of neighboring qudits (black dots) in a 2D lattice at various time substeps (layers). Each qudit is coupled to all of its neighbors at different substeps.  
   The half box at the boundary together with its neighbor on the opposite side combine to give a random unitary (identical matrix) acting on the corresponding pair of qudits, for periodic (open) boundary condition.
	We prove that, as in the 1D case (Fig.~\ref{fig:i}), the quasi-energy spectrum for this type of Floquet random quantum circuits with arbitrary ordering of the quantum gates exhibits universal statistical properties described by the CUE ensemble, in the limit of large onsite Hilbert space dimension $q\rightarrow \infty$, for both periodic and open boundary conditions. 
	This statement extends to arbitrary dimensions (see appendix~\ref{sec:AppFRC-HD}).
	}
\end{figure}

The spectral statistics of the Floquet operator of time-periodic systems can be extracted from this effective field theory.
In contrast to the aforementioned model-specific methods, the field-theoretic approach is applicable to a wide class of Floquet systems, and is particularly useful for random quantum circuits composed of local random gates.
To demonstrate the usefulness of this method, we apply it to study the spectral statistics of a family of Floquet random quantum circuits, some of which are shown in Figs.~\ref{fig:i} and~\ref{fig:ii}.
These Floquet circuits are composed of random unitary matrices which are drawn randomly and independently from the Circular Unitary Ensemble (CUE) and are applied to all pairs of neighboring qudits at various time substeps during one period. 
Among these circuits, the brickwork circuit depicted in Fig.~\ref{fig:i}(a) was studied earlier by Chan, De~Luca, and Chalker in Ref.~\cite{Chalker-1} (see also a related work Ref.~\cite{Chalker-4}), and the RMT spectral statistics were derived in the limit of a large on-site Hilbert space  dimension  $q\rightarrow \infty$, using a generalized diagrammatic approach initially constructed in Ref.~\cite{Beenakker} for the Weingarten calculus~\cite{Samuel, Weingarten, collins2003,collins2006,collins2021,Weingarten-rev,Zee}.
Applying the field theoretical approach, we rederive this result and also show that the universal RMT statistics is preserved under an arbitrary reordering of the two-qudit gates,
%(as long as different unitaries acting on the same qudit and its neigbors are applied at different substeps), 
for both periodic and open boundary conditions.
In particular, we show that the effective field theory describing the Floquet quantum circuit of this type is identical to that of the CUE ensemble.
Moreover, we consider higher-dimensional generalizations of this family of the Floquet circuits, and find that their quasi-energy spectra all exhibit the RMT statistics (see Fig.~\ref{fig:ii} for some of the 2D examples).

The field theoretical approach is not only useful  in the study of the spectral statistics of Floquet random quantum circuits, but can also be applied to the analysis of more generic properties of a wide class of random quantum circuits (which are not necessarily time periodic).
To show this, within the same sigma model framework, we rederive the known results for Weingarten calculus, which has extensive applications in studies of quantum circuits (see for example Refs.~\cite{ Nahum-2018,Sondhi,Nahum-2017,HJ-2018, Huse,Ludwig-1,Chalker-1}).
When considering a quantum circuit where all or part of the quantum gates are random unitaries drawn from some ensembles with Haar probability, 
one usually encounters polynomial functions of the matrix elements integrated over the relevant group with Haar measure (or equivalently averaged over the relevant ensemble). 
This type of Haar integrals can be computed using the Weingarten calculus~\cite{Samuel, Weingarten, collins2003,collins2006,collins2021,Weingarten-rev,Beenakker}.
In the present paper, we rederive the results for the Haar integrals of products of the matrix elements of a unitary matrix over the unitary group (i.e., moments of CUE random matrices) using the field theoretical approach. The derivation can be generalized to other compact Lie groups or compact symmetric spaces~\cite{collins2006,collins2009,Wg-COE,Wg-SymSpace}.

The rest of the paper is organized as follows:
In section~\ref{sec:CUE}, we introduce a standard diagnostic of the statistics of quasi-energies of time-periodic systems and briefly review the spectral statistical properties of the CUE ensemble.
In section~\ref{sec:sigma}, we present an effective field theory which is formulated as a replica sigma model and can be used to investigate the spectral statistics of a wide class of Floquet quantum systems.
This effective field theory is applied in section~\ref{sec:FRC} to study a family of Floquet random quantum circuits related to the brickwork circuit in Ref.~\cite{Chalker-1} by a reordering of the Haar random unitary gates (which couple the neighboring qudits) and by the higher-dimensional generalization. 
We show that the statistical properties of the quasi-energy spectra of the Floquet random quantum circuits of this type are universally described by the CUE ensemble, for both the periodic and open boundary conditions.
Using the same sigma model method, we rederive the results for the Weingarten calculus for the unitary group in section~\ref{sec:Wg}.
Finally, in section~\ref{sec:con} we conclude with a discussion of directions for future studies.
The appendices are devoted to technical details. 
In appendices~\ref{sec:AppFRC} and~\ref{sec:AppFRC-HD}, we provide the calculation of the second- and fourth- order moments of the Floquet operator for the family of Floquet random quantum circuits studied in the current paper,  in 1D and in higher dimensions, respectively. 
We prove that, for arbitrary ordering of the two-qudit unitary gates, and in any dimension, these moments are identical to that of the CUE ensemble.
In appendix~\ref{sec:nonint}, we study a noninteracting Floquet model and examine the higher oder fluctuations in the effective field theory. We show that the quartic order fluctuations of this noninteracting model give rise to a larger contribution to the level correlation function, compared with their chaotic counterparts, and are therefore no longer negligible.
In appendix~\ref{sec:Wg-expan}, we rederive the asymptotic behavior of the Weingarten function for a unitary group of dimension $q$ in the large $q\rightarrow \infty$ limit. 
Appendix~\ref{sec:Wg-Rec} contains a derivation of the recursion relation for the Weingarten function.

\section{Diagnostic of spectral statistics of time-periodic systems}~\label{sec:CUE}

For time-periodic (Floquet) systems, the statistics of the quasi-energy spectra serve as a diagnostic of quantum chaos. The quasi-energies $\left\lbrace  \theta_i \right\rbrace$ are the eigen-phases of the time evolution operator over one period $U$ ( i.e., the Floquet operator): $U\left|n\right\rangle=e^{i\theta_n}\left|n\right\rangle$.
The quasi-energy density can be expressed as 
\begin{align}\label{eq:rho}
\rho(\phi)
=
\sum_{i=1}^{N}\delta_{2\pi}(\phi-\theta_i),
%	=
%\frac{1}{2\pi}\sum_{i}\sum_{n=-\infty}^{\infty}e^{-in(\phi-\theta_i)}.
\end{align}
where $N$ is the Hilbert space dimension, and $\delta_{2\pi}(\phi)$ represents the $2\pi$-periodic delta function defined as $\delta_{2\pi}(\phi)=\sum_{n=-\infty}^{\infty}e^{-in\phi}/2\pi$.
For brevity, the subscript $2\pi$ will be omitted from now on.

For an ensemble of random Floquet systems, the two-point correlation function of the quasi-energy density is usually introduced as:
\begin{align}\label{eq:R2}
\begin{aligned}
R_2(\phi_1,\phi_2)
=&
\left\langle 
\rho(\phi_1) \rho(\phi_2)
\right\rangle,
\end{aligned}
\end{align}	
where the angular bracket denotes the ensemble averaging.
The correlation function, $R_2(\phi_1,\phi_2)$, measures the probability of finding two quasi-energies at $\phi_1$ and $\phi_2$, and 
is one of the widely used probes of the spectral statistics.
In the present paper, in addition to the ensemble averaging, we also perform the averaging over the entire quasi-energy spectrum:
\begin{align}
\begin{aligned}
\bar{R}_2(\Delta \phi)
=&
\int_0^{2\pi} \frac{d\phi_0}{2\pi} R_2(\phi_0+\Delta \phi/2,\phi_0-\Delta \phi/2).
%=
%\left\langle 
%\int_0^{2\pi} \frac{d\phi_0}{2\pi}
%\rho(\phi_0+\Delta \phi/2) \rho(\phi_0-\Delta \phi/2)
%\right\rangle .
%=
%\left\langle 
%\frac{1}{(2\pi)^2}\sum_{i,j=1}^{N}\sum_{n=-\infty}^{\infty}
%e^{-in(\Delta \phi-\theta_i+\theta_j)}
%\right\rangle,
\end{aligned}
\end{align}	
This averaging~\cite{Altland-rev} is similar to the energy integration employed in Refs.~\cite{Andreev-Ballistic,Andreev-Ballistic-2} to derive a ballistic sigma model for individual Hamiltonian systems, and allows us to extract information about the correlation function of an arbitrary pair of quasi-energy levels with the  separation $\Delta \phi$, irrespective of their positions in the quasi-energy spectrum on the Floquet circle (see also, Ref.~\cite{Zoller}).

The Fourier transform of the two-level correlation function
is known as the spectral form factor:
\begin{align}
    K(t)
	=
	\int_0^{2\pi}
	d\phi_1
	\int_0^{2\pi}
	d\phi_2
	R_2(\phi_1,\phi_2)
		e^{-i(\phi_1-\phi_2)t}
	=
    2\pi
	\int_0^{2\pi}
	d(\Delta \phi)\,
	\bar{R}_2(\Delta \phi)	e^{-i \Delta \phi\, t},
\end{align}	
which is equivalent to 
\begin{align}\label{eq:SFF}
	K(t)
	=
	\left\langle 
	\left| \Tr  \left( U^t \right) \right|^2 
	\right\rangle
	=
	\left\langle 
	\sum_{i,j=1}^{N}
	e^{-it(\theta_i-\theta_j)}
	\right\rangle.
\end{align}
From this definition, one can immediately see that $K(t)=K(-t)$ and $K(t=0)=N^2$. Moreover, assuming no degeneracy in the quasi-energy spectrum, in the large time limit $t\rightarrow \infty$, the off-diagonal ($i\neq j$) terms in the summation in Eq.~\eqref{eq:SFF} vanish upon ensemble averaging due to the random phases, and the SFF acquires the value of $K(t\rightarrow \infty)=N$ (plateau).

Chaotic Floquet systems without the time-reversal invariance are expected to exhibit the same spectral statistics properties as that of the Circular Unitary Ensemble (CUE)~\cite{Dyson-I,Haake,Mehta}, which is an ensemble of unitary matrices with Haar probability measure.
For the CUE ensemble, the spectrum is statistically homogeneous, and the mean level density is given by $\bar{\rho}=N/2\pi$. 
The two-level correlation function $R_2(\phi_1,\phi_2)$ depends only on the level separation $\Delta \phi=\phi_1-\phi_2$ and therefore is equal to its energy-averaged value~\cite{Mehta}:
\begin{align}\label{eq:R2-CUE}
\begin{aligned}
	R_2(\phi_1,\phi_2)
	=
	\bar{R}_2(\Delta \phi)
	=
	-\frac{1}{4\pi^2}
	 \dfrac{\sin^2 ({N\Delta \phi/2})}{\sin^2\left( \Delta \phi/2\right) }
	+
	\frac{N^2}{4\pi^2}
	+
	\frac{N}{2\pi} \delta (\Delta \phi).
\end{aligned}
\end{align}
At nonzero $\Delta\phi$, the first term in the equation above corresponds to the connected part of the two-level correlation function defined as
\begin{align}\label{eq:R2_con}
R^{\rm con}_2(\phi_1,\phi_2)
=
\left\langle 
\rho(\phi_1) \rho(\phi_2)
\right\rangle
-
\left\langle 
\rho(\phi_1) 
\right\rangle
\left\langle 
\rho(\phi_2)
\right\rangle.
\end{align}
In the large $N$ limit, after rescaling the quasi-energy $\varepsilon=\bar{\rho} \Delta \phi$ and keeping it finite, the connected two-level correlation function assumes the form identical to that of the Gaussian Unitary Ensemble (GUE)~\cite{Mehta,Haake}:
\begin{align}\label{eq:R2con-CUE}
\begin{aligned}
	\frac{1}{\bar{\rho}^2} R_2^{\rm con}(\Delta \phi)
	=
	-
	\left( \dfrac{\sin(\pi \varepsilon)}{\pi \varepsilon} \right)^2.
\end{aligned}
\end{align}
From Eq.~\eqref{eq:R2-CUE}, one finds that the SFF of the CUE ensemble acquires the form
\begin{align}\label{eq:SFF-CUE}
\begin{aligned}
K(t)
=
\min (|t|, N)
+N^2 \delta_{t,0}.
\end{aligned}
\end{align}
It exhibits a linear ramp until a plateau sets in at $t=N$.
This linear ramp reflects the repulsion between the quasi-energies and is expected to be a universal feature of the quantum chaotic systems with broken time-reversal symmetry.

\section{Replica sigma model for generic Floquet systems}\label{sec:sigma}

%\textcolor{blue}{
%In Ref.~\cite{Zirnbauer_1996}, Zirnbauer  constructed
%a supersymmetric sigma model using the color-flavor transformation~\cite{Zirnbauer_2021} and rederived the spectral properties of the CUE ensemble.  
%This sigma model is similar to the one formulated for disordered system~\cite{Wegner-1979,Efetov},
%but is particular useful to study systems modeled not by Hamiltonian but by scattering matrix or evolution operator~\cite{Zirnbauer_1998, Zirnbauer_1999}. 
%In particular, it has been used to investigate the dynamical localization in quantum kicked rotor~\cite{QKR,Tian_2010,Tian_2011,Tian_2012}, and  the eigenenergy and eigenfunction statistics of quantum graph~\cite{graph-1,graph-2,graph-eigenfun,graph-eigenfun-2}. 
%%%Quantum Hall ???
%See Ref.~\cite{Altland-rev} for a review.
%In the following, we reformulate this sigma model using replica instead of supersymmetric method, and apply it to Floquet quantum circuits.}

In this section, we reformulate the sigma model initially constructed in Refs.~\cite{Zirnbauer_1996,Zirnbauer_1998, Zirnbauer_1999,QKR} using the replica trick instead of the supersymmetric method, and obtain an ensemble averaged effective field theory from which one can extract information about the statistical properties of the quasi-energy spectra for generic time-periodic quantum systems. 

\subsection{Generating function for level correlation function}

Consider now an ensemble of Floquet systems whose Floquet operators $U$ are from an arbitrary ensemble of unitary matrices. 
Note that, for many-body systems, $U$ represents the many-body Floquet operator that acts in the many-body Hilbert space.
The statistical properties of the quasi-energies of this ensemble can be obtained from the following generating function using the replica trick
\begin{align}\label{eq:Z-0}
\begin{aligned}
\mathcal{Z}^{(R)}(\alpha,\beta)
=&
\left\langle 
\int_0^{2\pi} \frac{d\phi}{2\pi}
\left[ 
\det \left( 1-\alpha e^{i\phi} U\right) 
\det  \left( 1-\beta e^{-i\phi} U^{\dagger}\right) 
\right]^R 
\right\rangle.
\end{aligned}
\end{align}
Here $\alpha$ and $\beta$ are two complex numbers, and $R$ represents the replica number which will be set to zero (the replica limit) at the end of the calculation. 
Taking derivatives of the generating function $\mathcal{Z}^{(R)}(\alpha,\beta)$ with respect to the complex variables $\alpha$ and $\beta$,  multiplying the result by $\alpha\beta/R^2$ and then taking the replica limit $R\rightarrow 0$, we find
\begin{align}\label{eq:Cab}
\begin{aligned}
\mathcal{C}(\alpha,\beta)
\equiv&
\lim_{R\rightarrow 0} \frac{\alpha\beta}{R^2}
\dfrac{\partial^2\mathcal{Z}^{(R)}(\alpha,\beta)}{\partial \alpha \partial \beta}
=
\left\langle 
\int_0^{2\pi} \frac{d\phi}{2\pi}
\Tr 
\left(  \dfrac{\alpha e^{i\phi} U} {1-\alpha e^{i\phi} U} \right)
\Tr 
\left( \dfrac{\beta e^{-i\phi} U^{\dagger}}{1-\beta e^{-i\phi} U^{\dagger}}\right)
\right\rangle 
\\
=&
\left\langle 
\int_0^{2\pi} \frac{d\phi}{2\pi}
\sum_{n,m=1}^{\infty}
\Tr 
\left(  \alpha e^{i\phi} U\right)^n
\Tr 
\left( \beta e^{-i\phi} U^{\dagger}\right)^m
\right\rangle 
\\
=&
\sum_{n=1}^{\infty}
K(n)
(\alpha\beta)^n.
\end{aligned}
\end{align}
In this equation, the powers of $U$ and $U^{\dagger}$ (i.e., $n$ and $m$) need to be identical to have a nonvanishing contribution to $\mathcal{C}(\alpha,\beta)$ after the integration over $\phi$.

The two-level correlation function $\bar{R}_2(\Delta \phi)$ can be obtained from $\mathcal{C}(\alpha,\beta)$, which is a weighted summation of the SFF $K(n)$ (Eq.~\eqref{eq:SFF})  at discrete time $n$,  by setting $\alpha=\beta=e^{i\Delta \phi/2}$, 
\begin{align}\label{eq:R2-C}
\begin{aligned}
%	&2\re C(a=b=\exp(i\Delta \phi/2))
%	=
%	\sum_{n=-\infty}^{\infty}
%	K(n)\exp(in\Delta \phi)
%	-
%	N^2
%	=
%	\sum_{n=-\infty}^{\infty}
%	\left\langle \sum_{i,j=1}^{N}\exp \left[ -in(\theta_i-\theta_j-\Delta \phi) \right]\right\rangle 
%	-
%	N^2
%	\\
%	=&
%	\left( 2\pi \right)^2R_2(\Delta \phi)-N^2
%	\\
\bar{R}_2(\Delta \phi)
=
\frac{1}{(2\pi)^2}
\left[  2\re \mathcal{C}(\alpha=\beta=e^{i\Delta \phi/2})+N^2 \right].
\end{aligned}
\end{align}	
Here we have used the fact that $K(-n)=K(n)$ and $K(0)=N^2$, with $N$ being the dimension of the Hilbert space.
We note that the higher-order correlation functions of the quasi-energy density can be evaluated in an analogous manner making use of a similar generating function~\cite{Zirnbauer_1996}.

%\textcolor{blue}{In the present paper, we are interested in the smooth part of the two-point correlation function where the individual level's position is smoothed over.
%	Therefore, we introduce an imaginary part to $\Delta \phi$:
%	$
%	1/N^2 \ll \im \Delta \phi \ll 1/N
%	$.
%	The delta function in the definition of quasi-energy density is then smeared into a Lorentzian function of width $\im \Delta \phi $, and the resulting $R_2$ is a smoothed correlation function.}

\subsection{Replica sigma model for level correlation function}

The generating function defined in Eq.~\eqref{eq:Z-0} can be expressed as a Grassmann path integral:
\begin{align}\label{eq:Z-1}
\begin{aligned}
\mathcal{Z}^{(R)}(\alpha,\beta)
=&
\left\langle 
\int_0^{2\pi} \frac{d\phi}{2\pi}
\int \D(\bpsi,\psi)
\exp\left\lbrace 
-\bpsi^{+,u}_{i}\left( \delta_{ij}-\alpha e^{i\phi} U_{ij} \right) \psi^{+,u}_j
\right. \right. 
\\
& \left. \left. \qquad
-\bpsi^{-,u}_{i} \left( \delta_{ij}-\beta e^{-i\phi} U^{\dagger}_{ij} \right) \psi^{-,u}_j
\right\rbrace 
\right\rangle .
\end{aligned}
\end{align}
The Grassmann $\psi^{s,u}_{i}$ carries three different indices: $u=1,2,...,R$ is the replica index, $i=1,2,...,N$ labels the Hilbert space, and $s=+/-$ distinguishes the contributions from the forward ($U$) and backward ($U^{\dagger}$) evolution operators.
Throughout the paper, we employ the convention that repeated indices imply the summation.

Using the color flavor transformation~\cite{Zirnbauer_1996,Zirnbauer_1998,Zirnbauer_1999, Zirnbauer_2021}, the integration over the center phase $\phi$ in Eq.~\eqref{eq:Z-1} can be converted into an integration over $RN \times RN$ complex matrix field $Z$:~\footnote{Here we call matrix $Z$ a `field', but note that it does not have time dependence and carries indices in the Hilbert and replica spaces only.}
\begin{align}
\begin{aligned}
&\mathcal{Z}^{(R)}(\alpha,\beta)
=
c_0
\left\langle 
\int \D(\bpsi,\psi)
\exp\left\lbrace 
-\bpsi^{+,u}_{i}\psi^{+,u}_{i}
-\bpsi^{-,u}_{i} \psi^{-,u}_{i}
\right\rbrace 
\right. 
\\
&\times
\left. 
\int \D(Z,Z^{\dagger})
\det(1+Z^{\dagger}Z)^{-2RN-1}
\exp\left\lbrace 
\alpha\bpsi^{+,u}_{i} Z^{uv}_{ij} U^{\dagger}_{jj'}\psi^{-,v}_{j'}
-
\beta \bpsi^{-,u}_{i} (Z^{\dagger})^{uv}_{ij}
U_{jj'}\psi^{+,v}_{j'}
\right\rbrace 
\right\rangle 
\\
=&
c_0
\left\langle 
\int \D(Z,Z^{\dagger})
\det(1+Z^{\dagger}Z)^{-2RN-1}
\right. 
\\
&\qquad\times
\left. 
\int \D(\bpsi,\psi)
\exp\left\lbrace 
-
\begin{bmatrix}
\bpsi^{+} & \bpsi^{-}
\end{bmatrix}
\begin{bmatrix}
1 & -\alpha Z U^{\dagger} 
\\
\beta Z^{\dagger}U & 1
\end{bmatrix}
\begin{bmatrix}
\psi^{+} 
\\ 
\psi^{-}
\end{bmatrix}
\right\rbrace 
\right\rangle .
\end{aligned}
\end{align}
Here $Z^{uv}_{ij}$ is a complex matrix which carries indices in both the replica space (labeled by $u,v$) and the Hilbert space (labeled by $i,j$).
$c_0$ is an indefinite unessential normalization factor, which can be determined in the $R\rightarrow 0$ limit.

Performing the Gaussian integration over $\psi$, we arrive at a sigma model representation of the generating function
\begin{align}\label{eq:SZ}
\begin{aligned}
&\mathcal{Z}^{(R)}(\alpha,\beta)
=
c_0
\left\langle 
\int \D(Z,Z^{\dagger})
e^{-S[Z^{\dagger},Z]}
\right\rangle ,
\\
&S[Z^{\dagger},Z]
=
(2RN+1)\Tr\ln \left( 1+Z^{\dagger}Z \right) 
-\Tr\ln \left( 1+\alpha \beta Z U^{\dagger} Z^{\dagger} U\right) .
\end{aligned}
\end{align}
It is sometimes convenient to make the transformation
\begin{align}
\begin{aligned}
&Q=
T\Lambda T^{-1},
\qquad
T=
\begin{bmatrix}
1  & -Z
\\
Z^{\dagger}  & 1
\end{bmatrix},
\qquad
\Lambda
=
\begin{bmatrix}
1  & 0
\\
0 & -1
\end{bmatrix}
,
\end{aligned}
\end{align}
after which the action becomes
\begin{align}\label{eq:SQ}
\begin{aligned}
&S[Q]
=
-
2RN\Tr\ln 
\left( \frac{Q\Lambda+1}{2} \right) 
-
\Tr\ln 
\left[ 
\frac{1}{2}
\left( 
1
-
\begin{bmatrix}
\alpha U & 0
\\
0 & \beta U^{\dagger}
\end{bmatrix}
\right)
Q\Lambda
+
\frac{1}{2}
\left( 
1
+
\begin{bmatrix}
\alpha U & 0
\\
0 & \beta U^{\dagger}
\end{bmatrix}
\right)
\right] 
.
\end{aligned}
\end{align}
Similar to the sigma model for disordered systems~\cite{Wegner-1979,Efetov,Kamenev-book}, the matrix field $Q$ stays on the manifold with the constraints $\Tr Q=0$ and $Q^2=1$.
Eq.~\eqref{eq:SZ} (or equivalently Eq.~\eqref{eq:SQ}) is a replica version of the supersymmetric sigma model  derived earlier to study the spectral statistics of the circular ensembles and quantum chaotic maps~\cite{Zirnbauer_1996,Zirnbauer_1998,Zirnbauer_1999,Altland-rev}.

\subsection{Ensemble averaged effective theory}

Starting from Eq.~\eqref{eq:SZ}, we then perform the ensemble averaging and derive an effective field theory for the matrix field $Z$ for any ensemble of Floquet systems. Note that Eq.~\eqref{eq:SZ} can be rewritten as
\begin{align}\label{eq:Seff-0}
\begin{aligned}
\mathcal{Z}^{(R)}(\alpha,\beta)
=\,&
c_0
\int \D(Z,Z^{\dagger})
e^{-S_{\rm eff}[Z^{\dagger},Z]},
\\
S_{\rm eff}[Z^{\dagger},Z]
\equiv \,&
(2RN+1)\Tr\ln \left( 1+Z^{\dagger}Z \right) 
-\ln
\left\langle 
\exp \left[ \Tr\ln \left( 1+\alpha \beta Z U^{\dagger} Z^{\dagger} U\right)  \right] 
\right\rangle.
\end{aligned}
\end{align}
Until now, no approximations have been made and the expression above is exact.
However, it is difficult to perform the ensemble averaging in the second term in the action $S_{\rm eff}$, especially for many-body systems whose Floquet operator has a complicated structure in the many-body Hilbert space.
To proceed, we expand $S_{\rm eff}$ around the saddle point $Z=0$ in powers of fluctuation $Z$ and carry out the ensemble averaging term by term.
Up to quartic order in $Z$, the action $S_{\rm eff}$ is given by
\begin{subequations}\label{eq:Seff}
\begin{align}
	&
	S_{\rm eff}[Z^{\dagger},Z]
	=
	S_{\rm eff}^{(2)}[Z^{\dagger},Z]
	+
	S_{\rm eff}^{(4)}[Z^{\dagger},Z]
	+
	O(Z^6),
	\\
	&
	\begin{aligned}\label{eq:Seff2}
	S_{\rm eff}^{(2)}[Z^{\dagger},Z]
	= &
	\sum_{i_1,...,i_4}
	\tr \left( Z^{\dagger}_{i_1 i_2}  Z_{i_3 i_4}  \right) 
	\left[ 
	(2RN+1) \delta_{i_2,i_3} \delta_{i_4,i_1}
	-
	\alpha\beta
	\left\langle U_{i_2 i_3} U^{\dagger}_{i_4 i_1} \right\rangle
	\right]  ,
	\end{aligned}
	\\
	&\begin{aligned}\label{eq:Seff4}
	 S_{\rm eff}^{(4)}[Z^{\dagger},Z]
%	\equiv 
%	-\frac{(\textcolor{blue}{2RN+1})}{2}
%	\Tr \left( (Z^{\dagger}Z)^2 \right) 
%	+
%	\frac{(\alpha\beta)^2}{2}
%	\left\langle \Tr\left((Z^{\dagger} UZ U^{\dagger})^2\right)\right\rangle 
%	\\
%	&
%	-
%	\frac{(\alpha\beta)^2}{2}
%	\left\langle \Tr^2 \left(Z^{\dagger} UZ U^{\dagger}\right)\right\rangle
%	+
%	\frac{(\alpha\beta)^2}{2}
%	\left\langle \Tr \left(Z^{\dagger} UZ U^{\dagger}\right)\right\rangle^2 
%	\\
	=&
	-
	\frac{(\alpha\beta)^2}{2} 
	\sum_{i_1,...,i_8}
	\tr \left(Z^{\dagger}_{i_1 i_2} Z_{i_3 i_4}  \right)
	\tr \left(Z^{\dagger}_{i_5 i_6}  Z_{i_7 i_8}  \right)
	\\
	&\times
	\left[ 
	\left\langle U_{i_2 i_3} U^{\dagger}_{i_4i_1} U_{i_6i_7} U^{\dagger}_{i_8 i_5}\right\rangle
	-
	\left\langle U_{i_2 i_3} U^{\dagger}_{i_4 i_1} \right\rangle
	\left\langle U_{i_6 i_7} U^{\dagger}_{i_8 i_5}\right\rangle
	\right] 
	\\
	&+\frac{1}{2}\sum_{i_1,...,i_8}
	\tr \left(Z^{\dagger}_{i_1 i_2} Z_{i_3 i_4}  Z^{\dagger}_{i_5 i_6}  Z_{i_7 i_8} \right)
	\\
	&\times
	\left[ 
	-(2RN+1)
	\delta_{i_2,i_3} \delta_{i_4,i_5}\delta_{i_6,i_7}\delta_{i_8,i_1}
	+
	(\alpha\beta)^2 
	\left\langle U_{i_2 i_3} U^{\dagger}_{i_4 i_5} U_{i_6 i_7} U^{\dagger}_{i_8 i_1}\right\rangle
	\right] 
	.
	\end{aligned}
\end{align}	
\end{subequations}
In the present paper, we use `$\tr$'   to denote the trace operation that acts on the replica space only and `$\Tr$' to trace over both the replica and Hilbert spaces.
We note that the last term in $S_{\rm eff}^{(4)}$ (Eq.~\eqref{eq:Seff4}) gives rise to a contribution of higher order in the replica number $R$, compared with the remaining term in $S_{\rm eff}^{(4)}$, and is not important in the replica limit $R\rightarrow 0$.
%For the same reason, we can approximate $S_{\rm eff}$ in Eq.\eqref{eq:Seff-0} as
%\begin{align}
%\begin{aligned}
%S_{\rm eff}[Z^{\dagger},Z]
%\approx
%&
%(2RN+1)\Tr\left( Z^{\dagger}Z \right) 
%-
%\ln
%\left\langle 
%\exp \left[  \alpha \beta \Tr \left( Z U^{\dagger} Z^{\dagger} U \right)  \right] 
%\right\rangle.
%\end{aligned}
%\end{align}
%One can therefore see that, in an expansion in powers of $Z$,  the $2n$-order term in the effective action $S_{\rm eff}$ involves only the $2n$-th order cumulant for the Floquet operator $U$.

The specific forms of the moments of the Floquet operator in the effective theory are model dependent.
Once they are known, one can insert them into the expression for the effective action Eq.~\eqref{eq:Seff} to obtain an effective field theory which encodes the information about the  statistical properties of the quasi-energy spectrum. 
This field theoretical approach  is therefore applicable to a wide class of Floquet systems, including many-body and single-particle systems. One advantage of this method, compared with other model specific ones, is that it can be used to investigate why or when universal statistical behaviors emerge.
For Floquet random quantum circuits consisting of independent local unitary gates, the moments of the many-body Floquet operator are given by products of moments of the local unitaries, and therefore are usually not difficult to evaluate.
This makes the current field theoretical approach especially useful for random quantum circuits with local gates.

\section{Application to Floquet random quantum circuits}~\label{sec:FRC}

\subsection{Floquet random quantum circuits}

As an example, we apply the  effective field theory (Eq.~\eqref{eq:Seff}) derived in the previous section to study the spectral statistics of a family of Floquet quantum circuits composed of random local unitary gates. Figs.~\ref{fig:i} and~\ref{fig:ii} show the Floquet operators of some of these Floquet random quantum circuits, including the brickwork circuit (Fig.~\ref{fig:i}(a)) studied earlier in Ref.~\cite{Chalker-1}. The other two circuits depicted in panels (b) and (c) of Fig.~\ref{fig:i} can be obtained from the brickwork circuit by reordering the local gates.
Some examples of the 2D generalization of such Floquet circuits are depicted in Fig.~\ref{fig:ii}.
%Note that the brickwork quantum circuit depicted in Fig.~\ref{fig:i} (b) has been studied earlier by Chan et al~\cite{Chalker-1} using a generalized diagrammatic approach initially constructed in Ref.~\cite{Beenakker} for Weingarten calculus~\cite{collins2003,collins2006,collins2021,Weingarten-rev,Zee}. 
%Here we consider a family of similar Floquet quantum circuits which can be obtained from the brickwork circuit in Fig.~\ref{fig:i} (b) by reordering the local gates. 
We prove that Floquet random quantum circuits of this type, subject to either periodic or open boundary condition, are all described by the same effective field theory as that of the CUE ensemble in the limit of large on-site Hilbert space dimension, irrespective of the ordering of the local gates and the dimensionality of the lattice of qudits.

We first consider the Floquet quantum circuits consisting of a $1$D lattice chain of $L$ qudits, each of which contains $q \rightarrow \infty$ internal states. The dimension of many-body Hilbert space is $N=q^L$.
The time evolutions of these Floquet circuits are discrete and time periodic, and the evolutions over one period contain $2 \leq M\leq L$ substeps (where $M$ is model specific). 
For all of these Floquet quantum circuits, each qudit (labeled by an integer $n=1,2,...,L$) is coupled to its neighbors on the left hand side (at site $n-1$)  and the right hand side (at site $n+1$)  at two different substeps (labeled by integers $s^{(n-1,n)}$ and $s^{(n,n+1)}$, respectively). 
We consider both the periodic and open boundary conditions, and the qudit label $n$ is defined modulo $L$ for periodic boundary condition.
The local gate that couples a pair of neighboring qudits at sites $n$ and $n+1$ is given by a $q^2\times q^2$ random CUE matrix $w^{(n,n+1)}$, and is represented diagrammatically by a blue box in Fig.~\ref{fig:i}. Unitary gates acting on different pairs of neighboring qudits are independent and uncorrelated.
 We consider all possible orderings of these quantum gates represented by different configurations of $\left\lbrace 1 \leq  s^{(n,n+1)} \leq M | n=1,2,...,L' \right\rbrace$ with constraint
$
	s^{(n,n+1)} \neq s^{(n-1,n)}.
$
Here $L'$ is defined as $L'=L$ for periodic boundary condition and $L'=L-1$ for open boundary condition. 
The total number of substeps $M$ is given by the total number of different integers in the set  $\left\lbrace s^{(n,n+1)}  \right\rbrace$, and we consider all possible values of $2 \leq M\leq L'$.
For  the staircase circuit in Fig.~\ref{fig:i}(c), $M=L'$ and $s^{(n,n+1)}=n$, while for the brickwork circuit in Fig.~\ref{fig:i}(a), $M=2$ and $s^{(n,n+1)}=2-(n \Mod  2)$ (for open boundary condition or periodic boundary condition with even $L$).

For any of the Floquet random quantum circuits described above, the Floquet operator can be expressed as
\begin{align}\label{eq:U-S}
\begin{aligned}
&U=
W^{(\sigma(1),\sigma(1)+1)} W^{(\sigma(2),\sigma(2)+1)}
...
 W^{(\sigma(L'-1),\sigma(L'-1)+1)}W^{(\sigma(L'),\sigma(L')+1)} ,
\end{aligned}
\end{align}
where $\sigma \in S_{L'}$ represents a permutation of numbers $1,2,...,L'$.
$W$ is defined as
\begin{align}
	W^{(n,n+1)} =w^{(n,n+1)} \otimes 1^{(n,n+1)}.
\end{align}
$1^{(n,n+1)}$ represents an identical matrix operating in the Hilbert space of all sites except for $n$ and $n+1$. As mentioned earlier, $w^{(n,n+1)}$ acts on the qudits at sites $n$ and $n+1$, and is drawn randomly and independently from the CUE ensemble of dimension $q^2$.
We consider all possible quantum circuits whose Floquet operator can be expressed in the form of Eq.~\eqref{eq:U-S} for arbitrary permutation $\sigma \in S_{L'}$.
For example,  $\sigma$ is the identical permutation for the staircase circuit in Fig.~\ref{fig:i}(c), while for the brickwork circuit in Fig.~\ref{fig:i}(a) with periodic boundary condition and even $L$, $\sigma$ is given by
\begin{align}
\sigma(2k-1)
=
k,
\qquad
\sigma(2k)
=
L/2+k,
\qquad
1 \leq k\leq L/2.
\end{align}

In appendix~\ref{sec:AppFRC}, we prove that the Floquet operator $U$ given by Eq.~\eqref{eq:U-S} with arbitrary permutation $\sigma \in S_{L'}$  obeys the conditions:
\begin{subequations}\label{eq:cond}
	\begin{align}
	&\begin{aligned}\label{eq:cond-2}
	\left\langle 
	U_{\ib\jb} U^{\dagger}_{\jb'\ib'}
	\right\rangle 
	=
	\frac{1}{N}
	\delta_{\ib\ib'}\delta_{\jb\jb'},
	\end{aligned}
	\\
	&\begin{aligned}\label{eq:cond-4}
	\left\langle 
	U_{\ib_1\jb_1} U_{\ib_2\jb_2} U^{\dagger}_{\jb'_1\ib'_1} U^{\dagger}_{\jb'_2\ib'_2}
	\right\rangle 
	=
	\frac{1}{N^2}
	\left( 
	\delta_{\ib_1\ib_1'}\delta_{\jb_1\jb_1'}
	\delta_{\ib_2\ib_2'}\delta_{\jb_2\jb_2'}
	+
	\delta_{\ib_1\ib_2'}\delta_{\jb_1\jb_2'}
	\delta_{\ib_2\ib_1'}\delta_{\jb_2\jb_1'}
	\right)  .
	\end{aligned}
	\end{align}
\end{subequations}
Here the $L$-dimensional vector $\vex{i}=(i^{(1)},i^{(2)},...,i^{(L)})$ labels the many-body state of the circuits, and its $n$-th component $i^{(n)}=1,2,...,q$ indexes the single-particle state of the $n$-th qudit.
We note that Eq.~\eqref{eq:cond-2} holds for arbitrary $q$ while Eq.~\eqref{eq:cond-4}  is derived in the limit of  $q\rightarrow \infty$.

The above discussion can be straightforwardly generalized to higher-dimensional Floquet quantum circuits with similar configurations.
In particular, let us consider now a $D$-dimensional cubic lattice of qudits, with $L$ sites in each direction. The single-particle (many-body) Hilbert space dimension is $q\rightarrow \infty$ ($N=q^{L^D}$).
The time evolution is again discrete and periodic, and is composed of local two-qudit gates that couple separately all pairs of neighboring qudits.
During one period, each qudit is coupled to all of its neighboring qudits at different substeps by different local unitary gates drawn randomly and independently from the CUE ensemble of dimension $q^2$.
For this type of $D$-dimensional Floquet circuits with any possible ordering of these local gates, and with either periodic or open boundary condition,
we prove that the Floquet operator  $U$ still satisfies Eq.~\eqref{eq:cond}. 
In this case, $\vex{i}$ can be considered as a $L^D$ dimensional vector whose component $i^{(\vex{n})}=1,2,...,q$ labels the single-particle state of the qudit $\vex{n}$ in the $D$-dimensional lattice.
The derivation is relegated to appendix~\ref{sec:AppFRC-HD}.

Eqs.~\eqref{eq:cond-2} and~\eqref{eq:cond-4}  are also obeyed if $U$ is drawn randomly from a CUE ensemble of dimension $N\rightarrow \infty$. As a result,  the effective field theory (Eq.~\eqref{eq:Seff}) for the Floquet quantum circuits under consideration is equivalent to that of the CUE ensemble of dimension $N$, and
the two-level correlation functions for the current models are given by the CUE level correlation function in Eq.~\eqref{eq:R2-CUE}. 
%In the following, we provide the detailed derivation of the two-level correlation function $R_2(\Delta \phi)$ for the current models (or equivalently for the CUE ensemble).

%This family of higher-dimensional Floquet circuits is therefore also described by an effective field theory  identical to that of the CUE ensemble, and exhibits the CUE correlation function $R_2$ (Eq.~\eqref{eq:R2-CUE}).

\subsection{Quadratic fluctuations}\label{sec:R2FRC}

In the following, we  present the detailed derivation of the two-level correlation function $\bar{R}_2(\Delta \phi)$ for the quasi-energies of the Floquet random quantum circuits described above (or equivalently the CUE ensemble of the same dimension $N$) using the effective field theory Eq.~\eqref{eq:Seff} and the moments of the Floquet operator Eq.~\eqref{eq:cond}.

Substituting Eq.~\eqref{eq:cond-2} into the quadratic order effective action $S_{\rm eff}^{(2)}$ in Eq.~\eqref{eq:Seff2}, we obtain
\begin{align}\label{eq:Seff-2c}
\begin{aligned}
S_{\rm eff}^{(2)}
=&
(2RN+1)
\sum_{\ib_1,\ib_2}
\tr\left( Z^{\dagger}_{\ib_1 \ib_2}Z_{\ib_2 \ib_1} \right) 
-
\frac{\alpha\beta}{N}
\sum_{\ib_1,\ib_2}
\tr \left( Z^{\dagger}_{\ib_1 \ib_1}  Z_{\ib_2 \ib_2}  \right) .
%\\
%=
%&
%(\textcolor{blue}{2RN+1})
%\frac{1}{N} \sum_{k}
%\tr \left( X^{\dagger}(k) X(k)\right) 
%-
%\frac{\alpha\beta}{N}
%\tr \left( X^{\dagger}(0)  X(0)  \right)
%+
%(\textcolor{blue}{2RN+1})
%\sum_{\ib_1 \neq \ib_2}
%\tr\left( Y^{\dagger}_{\ib_1 \ib_2}Y_{\ib_2 \ib_1} \right) 
\end{aligned}
\end{align}
 We now divide $Z$ into the diagonal component $X_{\ib\ib}\equiv Z_{\ib\ib}$ and the off-diagonal component $Y_{\ib\jb}\equiv Z_{\ib\jb}$ (for $\ib\neq \jb$) in the Hilbert space, and Fourier transform the diagonal component $X_{\jb\jb}$ with respect to $\jb$:
\begin{align}\label{eq:FT}
\begin{aligned}
&X(\kb)
=
\sum_{\jb}
X_{\jb\jb} e^{-i2\pi \kb  \cdot \jb /q}.
%\qquad
%X^{\dagger}(k)
%=
%\sum_{j=1}^{N}
%X^{\dagger}_{jj} e^{i2\pi k j/N},
%\\
%X_{jj}
%=
%\frac{1}{N}
%\sum_{k=0}^{N-1}
%X(k) e^{i2\pi k j /N},
%\qquad
%X^{\dagger}_{jj}
%=
%\frac{1}{N}
%\sum_{k=0}^{N-1}
%X^{\dagger}(k) e^{-i2\pi k j/N}
\end{aligned}
\end{align}
Here $\kb$ is an $L^{D}$ dimensional vector and the summation over each component runs over $k^{(\vex{n})}=0,1,...,q-1$.
In terms of $X(\kb)$ and $Y_{\ib\jb}$, the effective action $S_{\rm eff}^{(2)}$ can be rewritten as
\begin{align}\label{eq:Seff-2c-XY}
\begin{aligned}
S_{\rm eff}^{(2)}
=\,
&
(2RN+1)
\frac{1}{N} \sum_{\kb}
\tr \left( X^{\dagger}(\kb) X(\kb)\right) 
-
\frac{\alpha\beta}{N}
\tr \left( X^{\dagger}(0)  X(0)  \right)
\\
&
+
(2RN+1)
\sum_{\ib_1 \neq \ib_2}
\tr\left( Y^{\dagger}_{\ib_1 \ib_2}Y_{\ib_2 \ib_1} \right).
\end{aligned}
\end{align}

From the equation above, one finds that the bare propagator for $X(\kb=0)$ acquires the form
\begin{align}\label{eq:Prop-X}
\begin{aligned}
	\left\langle X^{uv}(0) (X^{\dagger})^{v'u'}(0) \right\rangle_0 
	=\,
	\delta_{uu'}\delta_{vv'} 
	\dfrac{N}{(2RN+1)-\alpha\beta}.
\end{aligned}
\end{align}
Here the angular bracket with subscript $0$ represents the averaging over the Gaussian fluctuation of matrix $Z$ governed by the action $S_{\rm eff}^{(2)}$ (Eq.~\eqref{eq:Seff-2c-XY}).
Taking the replica limit $R \rightarrow 0$ and setting $\alpha=\beta$ to $e^{i\Delta \phi/2}$, the bare propagator for $X(0)$ becomes
\begin{align}\label{eq:Prop-X-r}
\begin{aligned}
\left\langle X^{uv}(0) (X^{\dagger})^{v'u'}(0) \right\rangle_0 
=\,
\delta_{uu'}\delta_{vv'} 
\dfrac{N}{1-e^{i\Delta \phi}}
,
\end{aligned}
\end{align}
which diverges when $\Delta \phi \rightarrow 0$. The corresponding mode is massless.

By contrast, the propagators for $X(\kb\neq 0)$ and $Y_{\ib \jb}$ are given by, respectively
\begin{align}\label{eq:Prop-XY}
\begin{aligned}
\left\langle X^{uv}(\kb) (X^{\dagger})^{v'u'}(\kb') \right\rangle_0 
=\,&
\delta_{uu'}\delta_{vv'} \delta_{\kb,\kb'}
\dfrac{N}{2RN+1},
\\
\left\langle Y^{uv}_{\ib\jb} (Y^{\dagger})^{v'u'}_{\jb'\ib'} \right\rangle_0 
=\,&
\delta_{uu'}\delta_{vv'} \delta_{\ib\ib'}\delta_{\jb\jb'}\frac{1}{2RN+1}.
\end{aligned}
\end{align}
$X(\kb\neq 0)$ and $Y_{\ib \jb}$ are therefore massive, and give rise to a $\alpha$, $\beta$ independent contribution to the generating function $\mathcal{Z}^{(R)}$ at the quadratic order.

In summary, the $Z$ fluctuations can be divided into two categories: the massless fluctuation $X(0)$ and the massive fluctuations $X(\kb\neq 0)$ and $Y_{\ib \jb}$.
The massive modes $X(\kb\neq 0)$ and $Y_{\ib \jb}$ contribute a nonessential constant to the generating function, while the soft mode $X(0)$  governs the spectral statistics~\cite{PRL,Winer}.

\subsection{Quartic fluctuations}

We now investigate the contribution to the self-energy from the quartic order fluctuations (Eq.~\eqref{eq:Seff4}).
Note that the last term in $S_{\rm eff}^{(4)}$ (Eq.~\eqref{eq:Seff4}) does not contribute in the replica limit $R \rightarrow 0$ due to its special structure in the replica space.
We can therefore focus on the first term in $S_{\rm eff}^{(4)}$ which can be expressed in terms of $X(\kb)$ and $Y$ as
\begin{align}\label{eq:Seff-4c-XY}
\begin{aligned}
&S_{\rm eff}^{(4-1)}
=
-\frac{(\alpha\beta)^2}{2N^{4}} 
\sum_{\kb_1,\kb_2}
\tr \left(X^{\dagger}(\kb_1) X (\kb_2)  \right)
\tr \left(X^{\dagger}(-\kb_1) X (-\kb_2)  \right)
\\
&
-
\frac{(\alpha\beta)^2}{2N^{2}} 
\sum_{\ib_1\neq \ib_2,\ib_3\neq \ib_4}
\tr \left(Y^{\dagger}_{\ib_1\ib_2} Y_{\ib_3\ib_4}  \right)
\tr \left(Y^{\dagger}_{\ib_2\ib_1}  Y_{\ib_4\ib_3}  \right)
\\
&
-
\frac{(\alpha\beta)^2}{2N^{3}}
\sum_{\ib_1\neq\ib_2}\sum_{\kb,\kb'}
\left[ 
\tr \left(Y^{\dagger}_{\ib_1\ib_2} X(\kb) \right)
\tr \left(Y^{\dagger}_{\ib_2\ib_1} X (-\kb) \right)
+
\tr \left(X^{\dagger}(-\kb) Y_{\ib_1\ib_2}  \right)
\tr \left(X^{\dagger}(\kb) Y_{\ib_2\ib_1}  \right)
\right]
.
\end{aligned}
\end{align}
Here we have used  Eq.~\eqref{eq:cond-4}.
Comparing with the quadratic action $S_{\rm eff}^{(2)}$  in Eq.~\eqref{eq:Seff-2c-XY}, one can see that the quartic action $S_{\rm eff}^{(4-1)}$ is of higher order in $1/N$. 
Note that this counting does not hold for nonergodic circuits and the higher order fluctuations become important (see appendix~\ref{sec:nonint}).

The self-energies for the $X$ and $Y$ components from the quartic interactions are given by, respectively,
	\begin{align}\label{eq:Sigma-c}
	\begin{aligned}
	(\Sigma_X)^{ab,ba}(\kb,\kb)
	=&
	\frac{(\alpha\beta)^2}{N^{4}} 
	\left\langle 
	X^{ab} (-\kb) 
	(X^{\dagger})^{ba} (-\kb) 
	\right\rangle_0 
	=
	\frac{(\alpha\beta)^2}{N^{3}} 
	\frac{1}{1-\alpha\beta \delta_{\kb,0}},
	\\
	(\Sigma_Y)^{ab,ba}_{\ib_1\ib_2,\ib_2\ib_1}
	=&
	\frac{(\alpha\beta)^2}{N^{2}} 
	\left\langle 
	Y_{\ib_2\ib_1}^{ab}  
	(Y^{\dagger})^{ba}_{\ib_1\ib_2}
	\right\rangle_0 
	=
	\frac{(\alpha\beta)^2}{N^{2}}.
	\end{aligned}
	\end{align}
Here we have taken the replica limit and ignored the self-energy corrections that vanish in the limit $R\rightarrow 0$.
From the equation above, we can see that the self-energy from the quartic order fluctuations is negligible in the large $N$ limit for the massive modes $X(\kb\neq 0)$ and $Y$. For the massless mode $X(0)$, the self-energy from the quartic interactions can be ignored if we consider a not too small energy separation $\Delta \phi \gg 1/N$. To recover the fine structure of the nearby quasi-energy levels, higher order fluctuations of $X(0)$ are needed.

We note that, for integrable systems, fluctuations beyond the quadratic order are no longer negligible, even when the quasi-energy separation $\Delta \phi$ being probed is much larger than the mean level spacing.
In appendix~\ref{sec:nonint}, we consider a noninteracting Floquet model whose single-particle dynamics within one period is generated by random CUE matrices.
In particular, the single-particle Floquet operator for each particle is independently drawn from the CUE ensemble. 
%This can be considered as a model for a time-periodic system of noninteracting particles, each of which exhibits independent chaotic dynamics.
We find that the action for the quadratic fluctuations of this noninteracting model is identical to that of the Floquet random quantum circuits considered in this section (or equivalently the CUE ensemble), and is given by Eq.~\eqref{eq:Seff-2c-XY}. However, the quartic fluctuations are governed by a different action which, compared with its chaotic counterpart Eq.~\eqref{eq:Seff-4c-XY}, leads to a much larger contribution to the self-energy of the $Z$ matrix field.
Unlike the chaotic model, the higher order fluctuations become important for this noninteracting model.

\subsection{Two-level correlation function}

For the chaotic Floquet random quantum circuits under consideration here, we can focus on the quadratic fluctuation's contribution and neglect the higher order corrections. 
$\mathcal{C}(\alpha,\beta)$ defined in Eq.~\eqref{eq:Cab} is then approximately given by
\begin{align}\label{eq:R2-2}
\begin{aligned}
\mathcal{C}(\alpha,\beta)
%\equiv&
%\lim_{R\rightarrow 0} \frac{\alpha\beta}{R^2}
%\dfrac{\partial^2\mathcal{Z}^{(R)}(\alpha,\beta)}{\partial a \partial b}
%\\
%=&
%\lim_{R\rightarrow 0} \frac{\alpha\beta}{R^2}
%c_0 \int \D(Z,Z^{\dagger}) e^{-S_{\rm eff}^{(2)}[Z,Z^{\dagger}] }
%\left[ 
%-\frac{\partial^2}{\partial a \partial b} S_{\rm eff}^{(2)}
%+
%\frac{\partial S_{\rm eff}^{(2)}}{\partial a} 
%\frac{\partial S_{\rm eff}^{(2)}}{\partial b} 
%\right] 
%\\
=&
\lim_{R\rightarrow 0} \frac{\alpha\beta}{R^2}
c_1
\left\langle 
\frac{1}{N}
\tr \left(X^{\dagger}(0)X(0)\right) 
+
\frac{\alpha\beta}{N^2}
\tr^2 \left(X^{\dagger}(0)X(0)\right) 
\right\rangle_0 
%\\
%=&
%\lim_{R\rightarrow 0} \frac{\alpha\beta}{R^2}
%c_1
%\left[ 
%\frac{1}{N} \frac{N}{2RN+1-\alpha\beta} R^2
%+
%\frac{\alpha\beta}{N^2}
%\left( \frac{N}{2RN+1-\alpha\beta}\right)^2 (R^4+R^2)
%\right] 
%\\
=
\frac{\alpha\beta}{(1-\alpha\beta)^2}.
\end{aligned}
\end{align}
Here the overall coefficient $c_1$ contains the unessential contribution from the massive modes and its value in the replica limit is determined from the fact that $\lim_{R\rightarrow 0}Z^{(R)}(\alpha,\beta)=1 $.
Using Eq.~\eqref{eq:R2-C}, we obtain the result for the two-level correlation function
\begin{align}\label{eq:R2-FRC}
\begin{aligned}
\bar{R}_2(\Delta \phi)
=
-\frac{1}{8\pi^2}
\dfrac{1}{\sin^2\left( \Delta \phi/2\right) }
+
\frac{N^2}{4\pi^2}.
\end{aligned}
\end{align}	
It is easy to see that  $\left\langle \Tr U \right\rangle=0$ for the Floquet circuits under consideration. Therefore the average quasi-energy density for any of these circuits is homogeneous and given by $\bar{\rho}=N/2\pi$. This also means that the first term in Eq.~\eqref{eq:R2-FRC} corresponds to the connected part of the two-level correlation function $R_2^{\rm con}(\Delta\phi\neq 0)$ defined in Eq.~\eqref{eq:R2_con}.
We emphasize that this equation applies to all of the Floquet quantum circuits under consideration.

Comparing Eq.~\eqref{eq:R2-FRC} with the exact CUE result Eq.~\eqref{eq:R2-CUE}, one can see that the smooth part of $\bar{R}_2(\Delta \phi)$ is recovered while an oscillatory term proportional to $\cos(N\Delta \phi)$ is missing.
To recover the oscillatory term or to extract the behavior of $\bar{R}_2(\Delta \phi)$ at small energy separation $\Delta \phi \lesssim 1/N$, nonperturbative information about higher order fluctuations %of the massless mode $X(0)$ 
is needed.
This may be obtained by consideration of nonstandard saddle points (see Ref.~\cite{Altland-rev}),
similar to the calculation of the level correlation function for the Hamiltonian systems described by the Gaussian ensembles~\cite{Andreev-Altshuler, Kamenev-GUE, Kamenev,Kamenev-Keldysh}.

\section{Weingarten calculus}\label{sec:Wg}

%\textcolor{blue}{
%The sigma model framework employed above not only is useful  in the study of spectral statistics of Floquet circuits, but can also be applied to the analysis of more-generic properties of a wide class of random quantum circuits (which are not necessarily time periodic).
%To show this, within this sigma model framework, we rederive the results of Weingarten calculus which has extensive applications in the quantum information theory, and in particular in studies of random quantum circuits (see for examples Refs.~\cite{collins2016, Nahum-2018,Sondhi,Nahum-2017,HJ2018, Huse,Ludwig-1,Chalker-1}).
%In particular,  when considering a quantum circuit where some/all of the quantum gates are random unitaries drawn from some ensembles with Haar probability, 
%one usually encounters polynomial functions of the matrix elements integrated over the relevant group with Haar measure (or equivalently averaged over the relevant ensemble). 
%This type of Haar integrals can be computed using the Weingarten calculus~\cite{Samuel, Weingarten, collins2003,collins2006,collins2021,Weingarten-rev,Beenakker}.
%In the current paper, we focus on the Haar integral over the unitary groups (i.e., average over the CUE ensemble), but the derivation can be generalized to other compact Lie groups or compact symmetric spaces~\cite{collins2006,collins2009,Wg-COE,Wg-SymSpace}. 
%}

To show that the field theoretical approach described above provides access not only to the spectral statistics but also to other generic properties of quantum circuits,
in this section, we rederive the  known results for the Weingarten calculus, which has been employed extensively in studies of quantum circuits.
Let us now consider a Haar integral of a product of the matrix elements of a unitary matrix $U$
over the unitary group in $q$ dimensions  $\mathcal{U}(q)$:
\begin{align}\label{eq:I}
\begin{aligned}
I
=&
\int_{\mathcal{U}(q)} 
dU
U_{i_1 j_1}U_{i_2 j_2}...U_{i_p j_p}
U^{\dagger}_{j_1' i_1'}U^{\dagger}_{j_2' i_2'}...U^{\dagger}_{j'_{p'} i'_{p'}}
\\
=&
\left\langle 
U_{i_1 j_1}U_{i_2 j_2}...U_{i_p j_p}
U^{\dagger}_{j_1' i_1'}U^{\dagger}_{j_2' i_2'}...U^{\dagger}_{j'_{p'} i'_{p'}}
\right\rangle_{\rm CUE},
\end{aligned}
\end{align}
where $dU$ denotes the normalized ($\int_{\mathcal{U}(q)} dU=1$)  Haar measure. 
This integral can be interpreted as the  product of the unitary matrix elements averaged over the CUE ensemble, denoted by the angular bracket with subscript CUE.
In this section and appendices~\ref{sec:Wg-expan} and~\ref{sec:Wg-Rec}, we use $U$ to denote a random CUE matrix (i.e., a Haar distributed random unitary matrix).

It has been found that this integral $I$ can be expressed in terms of the Weingarten function~\cite{Samuel, Weingarten, collins2003,collins2006,collins2021,Weingarten-rev,Beenakker}:
\begin{align}\label{eq:Wg}
\begin{aligned}
I
=
\delta_{p,p'}
\sum_{\tau,\sigma \in S_p}
\mathrm{Wg}(\tau^{-1}\sigma)
\prod_{k=1}^{p}
\delta_{i_k,i_{\sigma(k)}'}\delta_{j_k,j_{\tau(k)}'}
.
\end{aligned}
\end{align}
Here the double summation runs over all permutations $\tau,\sigma \in S_p$ of the integers $1,2,...,p$, and
the Weingarten function $\mathrm{Wg}(\tau^{-1}\sigma)$ depends only on the cycle structure of the product $\tau^{-1}\sigma$. More specifically,  the permutation $\tau^{-1}\sigma$ can be divided into $ m$ disjoint cycles:
\begin{align}\label{eq:cycle}
\begin{aligned}
\left( P_1^{(1)} \rightarrow P_2^{(1)}  \rightarrow ...\rightarrow P_{c_1}^{(1)} \right) 
\left( P_{1}^{(2)} \rightarrow P_{2}^{(2)} \rightarrow ...\rightarrow P_{c_2}^{(2)} \right) 
..
\left( P_{1}^{(m)} \rightarrow P_{2}^{(m)} \rightarrow ...\rightarrow P_{c_m}^{(m)}\right). 
\end{aligned}
\end{align}
Here $\left\lbrace P_{k}^{(l)} \right\rbrace $ are different integers from the set of integers under the permutation $\left\lbrace1,2,...,p \right\rbrace $, and satisfy $P_{k}^{(l)}=(\tau^{-1}\sigma)^{k-1} (P_{1}^{(l)})$ (for $k \leq  c_l$) and $P_{1}^{(l)}=(\tau^{-1}\sigma) P_{c_l}^{(l)}$.
The Weingarten function $\mathrm{Wg}(\tau^{-1}\sigma)$ depends only on the lengths of these disjoint cycles $\left\lbrace c_k| k=1,2,...,m\right\rbrace $, which obey the constraint $\sum_{k=1}^{m}c_k=p$, irrespective of their order.
In the following, we will denote the Weingarten function of the permutation $\tau^{-1}\sigma$ given by Eq.~\eqref{eq:cycle} as $\mathrm{Wg}(\tau^{-1}\sigma)=V^{(p)}_{c_1,c_2,...,c_m}$.
%For example, the permutation $\chi_0\in S_4$ defined as $(\chi_0(1),\chi_0(2),\chi_0(3),\chi_0(4))=(2,4,3,1)$ has the cycle structure $(1,2,4)(3)$, and its Weingarten function can be written as $\mathrm{Wg}(\chi_0)=V_{3,1}^{(4)}$.

The Weingarten function can be uniquely determined by the recursions relation~\cite{Samuel,Beenakker}: 
\begin{subequations}\label{eq:WgRec}
	\begin{align}
	&\begin{aligned}\label{eq:WgRec-a}
	qV_{c_1,...,c_m,1}^{(p+1)}
	+
	\sum_{s=1}^{m}
	c_s
	V_{c_1,...,c_{s-1},c_{s}+1,c_{s+1},...,c_m}^{(p+1)}
	=
	V_{c_1,...,c_m}^{(p)},
	\end{aligned}
	\\
	&\begin{aligned}\label{eq:WgRec-b}
	qV_{c_1,...,c_m}^{(p)}
	+
	\sum_{c=1}^{c_1-1}
	V_{c_1-c,c,c_2,...,c_m}^{(p)}	
	+
	\sum_{s=2}^{m}
	c_s
	V_{c_1+c_s,c_2,...,c_{s-1},c_{s+1},...,c_m}^{(p)}
	=
	0.
	\end{aligned}
	\end{align}
\end{subequations}
Here  $\left\lbrace c_i\right\rbrace$ represent an arbitrary set of $m$ positive integers that satisfy $\sum_{k=1}^{m}c_k=p$ in Eq.~\eqref{eq:WgRec-a}, and in Eq.~\eqref{eq:WgRec-b} an additional constraint $c_1\geq 2$ is imposed.

In the large $q\rightarrow \infty$ limit~\cite{Beenakker}, the Weingarten function of the identical permutation $id$ is given by $\mathrm{Wg}(id)=V^{(p)}_{1,1,...,1}=q^{-p}+O(q^{-p-2})$.
For all the remaining  permutations, the Weingarten function $V^{(p)}_{c_1,c_2,...,c_m}=O(q^{m-2p})$ is of higher order in $1/q$ (since the number of disjoint cycles $m<p$).
Using these results, one can easily see that Eq.~\eqref{eq:cond} is obeyed by the CUE random matrix $U$ of dimension $N$ in the large $N$ limit.

These results for Weingarten calculus have been obtained earlier using various methods~\cite{Weingarten-rev,Samuel,collins2003,collins2006,Beenakker,Zee}).
%For example, the Weingarten calculus has been used to study operator spreading~\cite{Nahum-2018,Sondhi}, entanglement growth~\cite{Nahum-2017}, measurement induced entanglement transitions~\cite{Ludwig-1}, localization~\cite{Huse}, chaos and thermalization~\cite{Chalker-1}
%in random or Floquet unitary circuits.
In the following, we rederive these results using a field theoretical method similar to the one employed earlier in the analysis of the statistics of quasi-energy spectra.
We believe that this field theoretical  approach can be straightforwardly generalized and serves as a general framework to study the quantum circuits. 

\subsection{Sigma model derivation for the Weingarten calculus}

To begin with, we rewrite the integral $I$ in Eq.~\eqref{eq:I} as a fermionic path integral
\begin{align}\label{eq:I0}
\begin{aligned}
I
=&
\int_{\mathcal{U}(q)} dU
\int \D(\bpsi,\psi)
e^{-S_U[\bpsi,\psi]}
F[\bpsi,\psi],
\\
S_U[\bpsi,\psi]
=&
\sum_{u=1}^{R'}
\sum_{l,k=1}^{q}
\left( 
\bpsi^{+,u}_k U_{kl} \psi^{+,u}_l
+
\bpsi^{-,u}_k U^{\dagger}_{kl} \psi^{-,u}_l
\right) ,
\\
	F[\bpsi,\psi]
=&
\prod_{k=1}^{p}
\left( \psi^{-,k}_{i_k}\bpsi^{-,k}_{j_k} \right) 
\prod_{l=1}^{p'}
\left( \psi^{+,l}_{j'_l}\bpsi^{+,l}_{i'_l} \right)
.
\end{aligned}
\end{align}
Here the fermionic field $\psi^{s,u}_{k}$ carries three indices that label, respectively, the component associated with $U/U^{\dagger}$  ($s=+/-$), the replica space ($u=1,2,...,R'$) and the Hilbert space in which the unitary matrix $U$ acts ($k=1,2,...,q$). Note here the replica number $R'$ is an integer given by $R' =\max(p,p')$, and, unlike in the earlier calculation of the level correlation function, it dose not need to be set to zero at the end.

In Eq.~\eqref{eq:I0}, we have made use of the Wick's theorem and the following identities
\begin{align}
\begin{aligned}
&
\left\langle 
\psi^{-,u}_{i_u}\bpsi^{-,v}_{j_v} 
\right\rangle_{S_U} 
=
\delta_{uv}
U_{i_u j_u}
,
\qquad	
\left\langle 
\psi^{+,u}_{j_u'}\bpsi^{+,v}_{i_v'} 
\right\rangle_{S_U} 
=
\delta_{uv}
U^{\dagger}_{j_u'i_u'},
\\
&
\left\langle 
\psi^{-}\bpsi^{+} 
\right\rangle_{S_U} 
=
\left\langle 
\psi^{+}\bpsi^{-} 
\right\rangle_{S_U} 
=
0.
\end{aligned}
\end{align}
The angular bracket with the subscript $S_U$ represents the functional averaging over the fermionic field $\psi$ with the weight $e^{-S_U[\bpsi,\psi]}$.
Note that in Eq.~\eqref{eq:I0} fermions with different replica  indices ($u$) or $U/U^{\dagger}$ indices ($s=\pm$) are uncoupled.
In fact, the replica space is introduced here such that, when computing the expectation value of $\left\langle F[\bpsi,\psi]\right\rangle_{S_U} $ using the Wick's contraction,
fermionic field $\psi^{\mp}$ with index $i_k$ ($j_k'$) has to pair with $\bpsi^{\mp}$ with index $j_k$ ($i_k'$) as they share the same replica index $k$, leading to the factor of $U_{i_k j_k}$ ($U_{j_k' i_k'}^{\dagger}$) in the integrand of $I$.

Applying the color-flavor transformation~\cite{Zirnbauer_1996, Zirnbauer_1998, Zirnbauer_1999, Zirnbauer_2021}, the Haar integral over the unitary matrix $U$ in Eq.~\eqref{eq:I0} can be converted into an integral over a complex $R' \times R'$ matrix field $Z$:
\begin{align}\label{eq:I1}
\begin{aligned}
I
=&
z_1
\int \D (Z,Z^{\dagger})
\det (1+Z^{\dagger}Z)^{-(2R'+q)}
\int \D(\bpsi,\psi)
e^{-S_Z[\bpsi,\psi]}
F[\bpsi,\psi], 
\\
=&
z_1
\int \D (Z,Z^{\dagger})
\det (1+Z^{\dagger}Z)^{-(2R'+q)}
\det (-ZZ^{\dagger})^q
\left\langle 
F[\bpsi,\psi]
\right\rangle_{S_Z} ,
\end{aligned}
\end{align}
where
\begin{align}
\begin{aligned}
S_Z[\bpsi,\psi]
=&
\sum_{u,v=1}^{R'}
\sum_{l=1}^{q}
\left( 
-
\bpsi^{+,u}_l Z_{uv} \psi^{-,v}_l
+
\bpsi^{-,u}_l Z^{\dagger}_{uv} \psi^{+,v}_l
\right) ,
\\
z_1^{-1}
=&
\int \D (Z,Z^{\dagger})
\det (1+Z^{\dagger}Z)^{-(2R'+q)}
\det(-ZZ^{\dagger})^q.
\end{aligned}
\end{align} 
Note that the matrix field $Z$ here acts  in the replica space only.
In the second equality of Eq.~\eqref{eq:I1}, we have integrated out the fermions governed by the new action $S_Z[\bpsi,\psi]$, and
the angular bracket with the subscript $S_Z$ represents the functional averaging  with the weight $\exp(-S_Z[\bpsi,\psi])$.
The normalization constant $z_1$ is determined from the fact that $I=1$  if we set $F=1$. 

After the color-flavor transformation, fermions with different replica indices interact through the matrix field $Z$, while those with different Hilbert space indices become uncoupled. The fermionic propagator now acquires the form
\begin{align}
\begin{aligned}
&
\left\langle 
\psi^{-,u}_{i_u}\bpsi^{+,v}_{i_v'} 
\right\rangle_{S_Z} 
=
-\delta_{i_u i_v'}
Z^{-1}_{uv}
,
\qquad
\left\langle 
\psi^{+,u}_{j_u'}\bpsi^{-,v}_{j_v} 
\right\rangle_{S_Z} 
=
\delta_{j_u' j_v}
(Z^{\dagger})^{-1}_{uv},
\\
&
\left\langle 
\psi^{-}\bpsi^{-} 
\right\rangle_{S_Z} 
=
\left\langle 
\psi^{+}\bpsi^{+} 
\right\rangle_{S_Z} 
=
0.
\end{aligned}
\end{align}
Making use of these results, we obtain
\begin{align}\label{eq:F}
\begin{aligned}
	\left\langle 
	F[\bpsi,\psi]
	\right\rangle_{S_Z}
	=&
	\delta_{p,p'}
	\sum_{\tau,\sigma \in S_p}
	\sgn (\tau^{-1}\sigma) 
	\prod_{k=1}^{p}
	\delta_{i_k,i_{\sigma(k)}'}\delta_{j_k,j_{\tau(k)}'}
	Z^{-1}_{k \sigma(k)}
	(Z^{\dagger})^{-1}_{\tau(k) k}.
\end{aligned}
\end{align}
Inserting Eq.~\eqref{eq:F} into Eq.~\eqref{eq:I1}, and applying the transformation
\begin{align}
\begin{aligned}
W = Z^{-1},
%\qquad
%W^{\dagger} = (Z^{\dagger})^{-1},
\end{aligned}
\end{align}
whose Jacobian leads to a contribution of  $2R'\tr \ln (WW^{\dagger}) $ to the action,
we find 
\begin{align}\label{eq:Wg0}
\begin{aligned}
&
I
=
\delta_{p,p'}
\sum_{\tau,\sigma \in S_p}
g(\sigma,\tau)
\prod_{k=1}^{p}
\delta_{i_k,i_{\sigma(k)}'}\delta_{j_k,j_{\tau(k)}'}
.
\end{aligned}
\end{align}
Here $g(\sigma,\tau)$ is given by
\begin{align}\label{eq:g}
\begin{aligned}
&g(\sigma,\tau)
=
\left\langle G[W,W^{\dagger} ;\sigma, \tau]\right\rangle_{W} 
\equiv
\dfrac{	
	\int \D (W,W^{\dagger})
	e^{-S_w[W,W^{\dagger}]}
	G[W,W^{\dagger};\sigma, \tau]
}
{\int \D (W,W^{\dagger})
	e^{-S_w[W,W^{\dagger}]}
},
\\
& S_w[W,W^{\dagger}]
=
	(2R'+q)\tr\ln \left( 1+WW^{\dagger} \right) 
,
\\
&G[W,W^{\dagger}; \sigma, \tau]
=
\sgn (\tau^{-1}\sigma) 
\prod_{k=1}^{p}W_{k \sigma(k)}
W^{\dagger}_{\tau(k) k}
.
\end{aligned}
\end{align}
The angular bracket with subscript $W$ represents the averaging over $W$ with the action $S_w[W,W^{\dagger}]$.

Using this field theoretical approach,  the Haar integral over the unitary matrix $U$, whose matrix elements are highly correlated due to the constraint $U U^{\dagger}=\bf{1}$, has been transformed to an integration over all complex matrix $W$ governed by the action $S_w$.
Note that Eq.~\eqref{eq:Wg0} is equivalent to the known result Eq.~\eqref{eq:Wg} if  $g(\sigma,\tau)=\mathrm{Wg}(\tau^{-1}\sigma )$. In the following, we're going to show that  $g(\sigma,\tau)$ defined by Eq.~\eqref{eq:g} depends only on the cycle structure of $\tau^{-1}\sigma $ and is indeed the Weingarten function $\mathrm{Wg}(\tau^{-1}\sigma )$.

\subsection{General properties of the Weingarten function}\label{sec:Wg-2}

If we apply the following transformation in Eq.~\eqref{eq:g}
\begin{align}
\begin{aligned}
&W_{ij}  \rightarrow  W_{i \chi(j)} ,
\qquad
W^{\dagger}_{ij} \rightarrow W^{\dagger}_{\chi(i)j} ,
\end{aligned}
\end{align}
where $\chi\in S_p$ represents an arbitrary permutation of $p$ numbers,
the action $S_w[W,W^{\dagger}]$ remains invariant, while $G[W,W^{\dagger}; \sigma, \tau]$ transforms to $G[W,W^{\dagger}; \chi\sigma, \chi\tau]$.
This proves that $g(\sigma,\tau)$ obeys the condition 
\begin{align}\label{eq:WG-1}
g(\sigma,\tau)=g(\chi\sigma,\chi\tau),
\end{align}
for an arbitrary permutation $\chi \in S_p$.
Setting $\chi$ to $\tau^{-1}$, one can immediately see that $g$ is a function of $\tau^{-1}\sigma$  only.

In an analogous way, one can also prove that $g(\sigma,\tau)$ depends only on the cycle structure  of $\tau^{-1}\sigma$:
\begin{align}\label{eq:V}
g(\tau,\sigma)=V^{(p)}_{c_1,c_2,...,c_m},
\end{align}
with $m$ being the total number of disjoint cycles in $\tau^{-1}\sigma$ and $c_i$ the length of the $i$-th cycle.
Rearranging the $W$ and $W^{\dagger}$ terms in $G[W,W^{\dagger};\sigma,\tau]$, we can rewrite $g(\sigma,\tau)$ as
\begin{align}\label{eq:g1}
\begin{aligned}
g(\sigma,\tau)
=&
\sgn (\tau^{-1}\sigma) 
\left\langle 
\prod_{k=1}^{m}
\left( 
\prod_{l=1}^{c_k}
W_{P_{l}^{(k)} \sigma(P_{l}^{(k)})}
W^{\dagger}_{\sigma(P_{l}^{(k)}) P_{l+1}^{(k)}}
\right) 
\right\rangle_{W} 
\\
=g(\tau^{-1}\sigma,id)
=&
(-1)^{p-m}
\left\langle 
\prod_{k=1}^{m}
\left( 
\prod_{l=1}^{c_k}
W_{P_{l}^{(k)} P_{l+1}^{(k)}}
W^{\dagger}_{P_{l+1}^{(k)} P_{l+1}^{(k)}}
\right) 
\right\rangle_{W} .
\end{aligned}
\end{align}
Here the integers $\left\lbrace P_l^{(k)}\right\rbrace $ denote the cycle structure of $\tau^{-1}\sigma$  (Eq.~\eqref{eq:cycle}) and they satisfy
$P_{l}^{(k)}=(\tau^{-1}\sigma)^{l-1} (P_{1}^{(k)})$. We introduced here the notation that  $P_{l}^{(k)}=P_{(l \Mod c_k)}^{(k)}$.
In the second equality,
we have used $\sgn(\tau^{-1}\sigma)=(-1)^{p-m}$ as well as Eq.~\eqref{eq:WG-1}.

Note that the action $S_w[W,W^{\dagger}]$ is invariant under the transformation
\begin{align}
\begin{aligned}
&W_{ij}  \rightarrow  W_{\chi(i) \chi(j)} ,
\qquad
W^{\dagger}_{ij} \rightarrow W^{\dagger}_{\chi(i) \chi(j)},
\end{aligned}
\end{align}
for arbitrary $\chi\in S_p$.
We now apply this transformation and choose the permutation $\chi$ defined by
\begin{align}
\chi(P_l^{(k)})
=
b_k+l,
\end{align}
where $b_k=\sum_{h=1}^{k-1}c_h$ for $k \geq 2$ and $b_1=0$.
After this transformation, $g(\sigma,\tau)$  becomes
\begin{align}\label{eq:g-c}
\begin{aligned}
g(\sigma,\tau)
=&
(-1)^{p-m}
\left\langle 
\prod_{k=1}^{m}
\left( 
\prod_{l=1}^{c_k}
W_{b_k+l,b_k+(l+1 \Mod c_k) }
W^{\dagger}_{b_k+(l+1 \Mod c_k), b_k+(l+1 \Mod c_k)}
\right) 
\right\rangle_{W}.
\end{aligned}
\end{align}
This equation shows that $g(\sigma,\tau)$ depends only on the lengths $\left\lbrace c_i |i=1,...,m \right\rbrace $  of the disjoint cycles of $\tau^{-1}\sigma$.

Using these results, we prove in appendix~\ref{sec:Wg-Rec} that the function $g(\sigma,\tau)$  defined in Eq.~\eqref{eq:g} satisfies the recursion relation Eq.~\eqref{eq:WgRec} and therefore is given by the Weingarten function $g(\sigma,\tau)=\mathrm{Wg}(\tau^{-1}\sigma)$ for $p\leq q$. See Ref.~\cite{Samuel} for a discussion about extrapolating  the results for the case of $p \leq q$ to $p >q$.
In appendix~\ref{sec:Wg-expan}, we also provide the derivation for the asymptotic behavior of the Weingarten function $g(\sigma,\tau)$ in the large $q\rightarrow \infty$ limit.

\section{Conclusion}\label{sec:con}

In summary, we derive an effective field theory which can be employed to investigate the spectral statistics of the Floquet operator for a large class of Floquet quantum systems.
We apply it to a family of Floquet random quantum circuits whose Floquet operators are composed of two-qudit random unitaries acting on pairs of neighboring qudits in a $D$-dimensional lattice. 
Universal RMT statistics has been found in the limit of infinite local Hilbert space dimensions, irrespective of the ordering of local gates, dimensionality of the qudit lattice, and the choice of the boundary condition (open or periodic).
This field theoretical approach has also been used to rederive the known results for the Weingarten calculus, which is a method to compute Haar integrals of polynomial functions of the matrix elements.

The wide applicability of the Weingarten calculus in quantum circuit studies suggests that this field theoretical approach may be generalized to investigate other fundamental features of quantum many-body systems, which is a direction for future works.
In particular, the field theory  may be useful for the investigation of phase transition in quantum circuits, such as the measurement-induced phase transition~\cite{Skinner,Fisher-1,Fisher-2,potter} which arises from the competition between the unitary dynamics and the projective measurements. A renormalization group analysis of the field theory may be available to carefully examine this phase transition.
Moreover, a mapping has been found between the entanglement growth (or the operator spreading in the high dimensional case) in random quantum circuits and the classical statistical mechanics problem of the surface growth in the Kardar-Parisi-Zhang universality class~\cite{Nahum-2017,Nahum-2018}. The field theoretical approach may provide some insight into this mapping in a more generic setting.

In the present paper, we focus on the Weingarten calculus for the unitary group having in mind random quantum circuits whose local gates are given by the CUE random matrices.
The derivation can be immediately generalized to other compact groups~\cite{collins2006,collins2009,Wg-COE,Wg-SymSpace},
making use of the color-flavor transformation for the corresponding group~\cite{Zirnbauer_2021}.
This generalization may have direct applications to random quantum circuits with different symmetry classes~\cite{HJ-2018}.
Moreover, as a simple example, the effective field theory given by Eq.~\eqref{eq:Seff} is employed above to study the spectral statistics of Floquet quantum circuits whose local gates are drawn from the CUE ensemble.
In fact, it is applicable to quantum circuits with random gates drawn from various ensembles of unitary matrices.
Therefore it is a useful analytical tool to study the Thouless energy - the energy separation below which the RMT statistics appears, and also to look for the putative transitions between the ergodic phase and many-body localization phase in various Floquet random quantum circuit models~\cite{Chalker-2,Chalker-5,Chalker-6,Huse,Huse-2,Prosen-Local}.

Replica trick is used in the present paper to derive the sigma model for generic Floquet quantum systems, which was originally constructed in the supersymmetric formalism~\cite{Zirnbauer_1996,Zirnbauer_1998,Zirnbauer_1999,QKR}. One advantage of using the replica method is that many terms in the action for higher order fluctuations are irrelevant in the replica limit and therefore can be omitted.
However, unlike the supersymmetric calculation which allows for a nonperturbative analysis~\cite{Zirnbauer_1996,Efetov}, the current replica calculation is perturbative and can not recover the oscillating term in the level correlation function. Consideration of nonstandard saddle points~\cite{Andreev-Altshuler, Kamenev-GUE, Kamenev,Altland-rev} may recover the missing oscillating term and is left for a future study.

\acknowledgments
This work was supported by the U.S. Department of Energy,
Office of Science, Basic Energy Sciences under Award
No. DE-SC0001911. Y.L. acknowledges a post-doctoral
fellowship from the Simons Foundation ``Ultra-Quantum Matter'' Research Collaboration. 
 
%\paragraph{Note added.} This is also a good position for notes added
%after the paper has been written.

\appendix

\section{Derivation of the moments of Floquet operator}~\label{sec:AppFRC}

In this appendix, we prove that the Floquet operator  $U$ given by Eq.~\eqref{eq:U-S} with arbitrary permutation $\sigma \in S_{L'}$ follows the conditions Eq.~\eqref{eq:cond-2} and~\eqref{eq:cond-4}, in the limit $q\rightarrow \infty$.
In other words, the moments of the Floquet operator $U$ for any Floquet random quantum circuit that is related to the brickwork circuit studied in Ref.~\cite{Chalker-1} by reordering the local quantum gates within one period are given by Eq.~\eqref{eq:cond}.
%, which are also satisfied by the CUE matrix with dimension $N$.
We make use of the fact that the moments of the Floquet operator $U$ are given by the product of the moments of the independent random CUE matrices $w^{(n,n+1)}$, which also take the form of Eq.~\eqref{eq:cond} after the replacement $N \rightarrow q^2$  and $\ib\rightarrow (i^{(n)},i^{(n+1)})$. Here $q^2$ and $(i^{(n)},i^{(n+1)})$ are, respectively, the dimension and label of the two-qudit Hilbert space in which the CUE matrix $w^{(n,n+1)}$ operates.

For any Floquet operator $U$ which takes the form of Eq.~\eqref{eq:U-S}, we can express its second order moment $\left\langle  U_{\vex{i}\vex{j}} U^{\dagger}_{\vex{j}'\vex{i}'} \right\rangle $ as
\begin{align}\label{eq:UU-1}
\begin{aligned}
	\left\langle U_{\vex{i} \vex{j}}U^{\dagger}_{\vex{j}'\vex{i}'}\right\rangle 
	=&
	\sum_{\vex{k},\vex{k}'}
	\prod_{n=1}^{L'}
	\left\langle 
	w^{(n,n+1)}_{a^{(n)},b^{(n+1)}; c^{(n)} ,d^{(n+1)} }
	(w^{(n,n+1)})^{\dagger}_{c'^{(n)},d'^{(n+1)}; a'^{(n)} ,b'^{(n+1)} }
	\right\rangle 
	\\
	=&
	\sum_{\vex{k},\vex{k}'}
	\prod_{n=1}^{L'}
	\frac{1}{q^2}
	\delta_{a^{(n)},a'^{(n)}}
	\delta_{b^{(n+1)},b'^{(n+1)}}
	\delta_{c^{(n)},c'^{(n)}}
	\delta_{d^{(n+1)},d'^{(n+1)}}
	.
\end{aligned}
\end{align}
As earlier, vector $\vex{i}$ labels the many-body state of all the qudits in the circuit and its $n$-th element $i^{(n)}$ indexes the single-particle state of the $n$-th qudit.
$\vex{k}=(k^{(1)},...,k^{(L)})$ ($\vex{k}=(k^{(2)},...,k^{(L-1)})$) is a $L$-dimensional ($(L-2)$-dimensional)  vector for periodic (open) boundary condition. $k^{(n)}$ labels the single-particle state at site $n$ at an intermediate substep and one has to sum over all possible states $k^{(n)}=1,...,q$.
$a^{(n)} \, (b^{(n+1)})$ and $c^{(n)} \,(d^{(n+1)})$ are respectively the row and column indices the label the Hilbert space of the site $n$ ($n+1$) for the CUE random matrix $w^{(n,n+1)}_{a^{(n)},b^{(n+1)}; c^{(n)} ,d^{(n+1)} }$, and they are
defined as follows.
Note first that only the unitaries $w^{(n,n+1)}$ and $w^{(n-1,n)}$ act on the qudit at site $n$, and
$\left\lbrace a^{(n)} ,b^{(n)} ,c^{(n)} ,d^{(n)}\right\rbrace $ depend on the ordering of these two unitaries. 
In particular, if $\sigma^{-1}(n) <\sigma^{-1}(n-1)$ (i.e., unitary $w^{(n,n+1)}$ is applied before $w^{(n-1,n)}$), we have
\begin{align}\label{eq:abcd-1}
\begin{aligned}
	a^{(n)}=i^{(n)},
	\quad
	c^{(n)}=k^{(n)},
	\quad
	b^{(n)}=k^{(n)},
	\quad
	d^{(n)}=j^{(n)},
%	\\
%	a'^{(n)}=i'^{(n)},
%	\quad
%	c'^{(n)}=k'^{(n)},
%	\quad
%	b'^{(n)}=k'^{(n)},
%	\quad
%	d'^{(n)}=j'^{(n)},
\end{aligned}
\end{align}
and otherwise,
\begin{align}\label{eq:abcd-2}
\begin{aligned}
	a^{(n)}=k^{(n)},
	\quad
	c^{(n)}=j^{(n)},
	\quad
		b^{(n)}=i^{(n)},
	\quad
	d^{(n)}=k^{(n)}.
%	\\
%	a'^{(n)}=k'^{(n)},
%	\quad
%	c'^{(n)}=j'^{(n)}
%	\quad
%	b'^{(n)}=i'^{(n)},
%	\quad
%	d'^{(n)}=k'^{(n)}.
\end{aligned}
\end{align}
For periodic boundary condition, Eqs.~\eqref{eq:abcd-1} and~\eqref{eq:abcd-2} apply to arbitrary site $n$ which is defined modulo $L$.
By contrast, for open boundary condition, these two equations hold for $1<n<L$. 
At the boundaries $n=1, L$, we instead have
\begin{align}\label{eq:abcd-3}
	a^{(1)}=i^{(1)},
	\qquad
	c^{(1)}=j^{(1)},
	\qquad
	b^{(L)}=i^{(L)},
	\qquad
	d^{(L)}=j^{(L)}.
\end{align}
The row and column indices for $(w^{(n,n+1)})^{\dagger} $ are defined in the same way. $\left\lbrace a'^{(n)} ,b'^{(n)} ,c'^{(n)} ,d'^{(n)}\right\rbrace $ are related to $\left\lbrace i'^{(n)} ,j'^{(n)} ,k'^{(n)}\right\rbrace $ in exactly the same way as $\left\lbrace a^{(n)} ,b^{(n)} ,c^{(n)} ,d^{(n)}\right\rbrace $ are related to $\left\lbrace i^{(n)} ,j^{(n)} ,k^{(n)}\right\rbrace $.

For both the cases of $\sigma^{-1}(n) <\sigma^{-1}(n-1)$ and $\sigma^{-1}(n) >\sigma^{-1}(n-1)$,
$\left\lbrace a^{(n)} ,b^{(n)} ,c^{(n)} ,d^{(n)}\right\rbrace $ are related with $\left\lbrace i^{(n)} ,j^{(n)} ,k^{(n)} ,k^{(n)}\right\rbrace $ by a permutation, and the following identity
\begin{align}
\delta_{a^{(n)},a'^{(n)}}
\delta_{b^{(n)},b'^{(n)}}
\delta_{c^{(n)},c'^{(n)}}
\delta_{d^{(n)},d'^{(n)}}
=
\delta_{i^{(n)},i'^{(n)}}
\delta_{j^{(n)},j'^{(n)}}
\left( \delta_{k^{(n)},k'^{(n)}}\right)^2,
\end{align}
is always true for periodic boundary condition (and for open boundary condition when $1<n<L$).
Making use of this property in Eq.~\eqref{eq:UU-1},  we find that the second order moment of the Floquet operator for the periodic boundary condition is given by
\begin{align}\label{eq:UU-2}
\begin{aligned}
\left\langle U_{\vex{i} \vex{j}}U^{\dagger}_{\vex{j}'\vex{i}'}\right\rangle 
=&
\frac{1}{q^{2L}}
\sum_{\vex{k},\vex{k}'}
\prod_{n=1}^{L}
\delta_{i^{(n)},i'^{(n)}}
\delta_{j^{(n)},j'^{(n)}}
(\delta_{k^{(n)},k'^{(n)}})^2
=
\frac{1}{q^{L}}
\prod_{n=1}^{L}
\delta_{i^{(n)},i'^{(n)}}
\delta_{j^{(n)},j'^{(n)}}.
\end{aligned}
\end{align}
The second order moment of $U$ for the open boundary condition can be evaluated in a similar way:
\begin{align}\label{eq:UU-3}
\begin{aligned}
\left\langle U_{\vex{i} \vex{j}}U^{\dagger}_{\vex{j}'\vex{i}'}\right\rangle 
=&
\frac{1}{q^{2(L-1)}}
\delta_{i^{(1)},i'^{(1)}}
\delta_{j^{(1)},j'^{(1)}}
\delta_{i^{(L)},i'^{(L)}}
\delta_{j^{(L)},j'^{(L)}}
\sum_{\vex{k},\vex{k}'}
\prod_{n=2}^{L-1}
\delta_{i^{(n)},i'^{(n)}}
\delta_{j^{(n)},j'^{(n)}}
(\delta_{k^{(n)},k'^{(n)}})^2
\\
=&
\frac{1}{q^{L}}
\prod_{n=1}^{L}
\delta_{i^{(n)},i'^{(n)}}
\delta_{j^{(n)},j'^{(n)}}.
\end{aligned}
\end{align}

As a concrete example,  consider the staircase circuit depicted in Fig.~\ref{fig:i}(c) with periodic boundary condition. In this case,
$\left\lbrace a^{(n)} ,b^{(n)} ,c^{(n)} ,d^{(n)}\right\rbrace $ are given by Eq.~\eqref{eq:abcd-2}, and the second order moment for the Floquet operator $U$ is given by
\begin{align}
\begin{aligned}
	\left\langle  
	U_{\vex{i}\vex{j}} U^{\dagger}_{\vex{j}'\vex{i}'} 
	\right\rangle 
	=&
	\sum_{\vex{k},\vex{k'}}
	\left\langle 
	w^{(1,2)}_{i^{(1)},i^{(2)}; k^{(1)} ,k^{(2)} }
	(w^{(1,2)})^{\dagger}_{k'^{(1)} ,k'^{(2)}; i'^{(1)},i'^{(2)} }
	\right\rangle
	\\
	&\times 
	\prod_{n=2}^{L-1}
	\left\langle 
	w^{(n,n+1)}_{k^{(n)},i^{(n+1)}; j^{(n)} ,k^{(n+1)} }
	(w^{(n,n+1)})^{\dagger}_{j'^{(n)} ,k'^{(n+1)} ; k'^{(n)},i'^{(n+1)}}
	\right\rangle 
	\\
	&\times
	\left\langle 
	w^{(L,1)}_{k^{(L)},k^{(1)}; j^{(L)},j^{(1)} }
	(w^{(L,1)})^{\dagger}_{j'^{(L)} ,j'^{(1)} ; k'^{(L)},k'^{(1)}}
	\right\rangle 
	\\
	&=
	\frac{1}{q^{L}} 
	\delta_{\vex{i},\vex{i'}} \delta_{\vex{j},\vex{j'}}   .
\end{aligned}
\end{align}
Reordering the two-qudit quantum gates only changes the row and column labels of $w^{(n,n+1)}$ and $(w^{(n,n+1)})^{\dagger}$ in the equation above but does not affect the final result.
Note that this result for the second order moment $\left\langle  U_{\vex{i}\vex{j}} U^{\dagger}_{\vex{j}'\vex{i}'} \right\rangle $ is valid for arbitrary $q$.  

We now evaluate the fourth order moment of $U$ and consider the $q \rightarrow \infty$ limit for simplicity. It can be expressed as
 \begin{align}\label{eq:UUUU-1}
\begin{aligned}
&\left\langle  
U_{\vex{i}_1 \vex{j}_1} U_{\vex{i}_2 \vex{j}_2} U^{\dagger}_{\vex{j}'_1\vex{i}'_1} U^{\dagger}_{\vex{j}'_2\vex{i}'_2} 
\right\rangle 
=
\sum_{\vex{k}_1,\vex{k}_1',\vex{k}_2,\vex{k}_2'}
\prod_{n=1}^{L'}
\left\langle 
w^{(n,n+1)}_{a^{(n)}_1,b^{(n+1)}_1; c^{(n)}_1,d^{(n+1)}_1 }
w^{(n,n+1)}_{a^{(n)}_2,b^{(n+1)}_2; c^{(n)}_2 ,d^{(n+1)}_2 }
\right. 
\\
&\times
\left. 
(w^{(n,n+1)})^{\dagger}_{c'^{(n)}_1,d'^{(n+1)}_1; a'^{(n)}_1 ,b'^{(n+1)}_1 }
(w^{(n,n+1)})^{\dagger}_{c'^{(n)}_2,d'^{(n+1)}_2; a'^{(n)}_2 ,b'^{(n+1)}_2 }
\right\rangle 
\\
=&
\sum_{\vex{k}_1,\vex{k}_1',\vex{k}_2,\vex{k}_2'}
\prod_{n=1}^{L'}
\frac{1}{q^4}
\sum_{s^{(n)}=1}^{2}
\left(
\begin{aligned}
&
\delta_{s^{(n)},1}
\delta_{a^{(n)}_1,a'^{(n)}_1}
\delta_{b^{(n+1)}_1,b'^{(n+1)}_1}
\delta_{c^{(n)}_1,c'^{(n)}_1}
\delta_{d^{(n+1)}_1,d'^{(n+1)}_1}
\\
&\times
\delta_{a^{(n)}_2,a'^{(n)}_2}
\delta_{b^{(n+1)}_2,b'^{(n+1)}_2}
\delta_{c^{(n)}_2,c'^{(n)}_2}
\delta_{d^{(n+1)}_2,d'^{(n+1)}_2}
\\
+&\delta_{s^{(n)},2}
\delta_{a^{(n)}_1,a'^{(n)}_2}
\delta_{b^{(n+1)}_1,b'^{(n+1)}_2}
\delta_{c^{(n)}_1,c'^{(n)}_2}
\delta_{d^{(n+1)}_1,d'^{(n+1)}_2}
\\
&\times
\delta_{a^{(n)}_2,a'^{(n)}_1}
\delta_{b^{(n+1)}_2,b'^{(n+1)}_1}
\delta_{c^{(n)}_2,c'^{(n)}_1}
\delta_{d^{(n+1)}_2,d'^{(n+1)}_1}
\end{aligned}
\right) .
\end{aligned}
\end{align}
Here $\vex{k}_{1,2}$ and $\vex{k}'_{1,2}$, similar to $\kb$ and $\kb'$, are $L$-dimensional ($(L-2)$-dimensional)  vectors whose components label the intermediate single-particle states for all sites (sites in the bulk of the lattice $1<n<L$) for periodic (open) boundary condition.
$\left\lbrace a^{(n)}_{1/2}, b^{(n)}_{1/2}, c^{(n)}_{1/2}, d^{(n)}_{1/2}\right\rbrace $ are related to $\left\lbrace i^{(n)}_{1/2} ,j^{(n)}_{1/2} ,k^{(n)}_{1/2}\right\rbrace $ in the same way as $\left\lbrace a^{(n)} ,b^{(n)} ,c^{(n)} ,d^{(n)}\right\rbrace $ are related to $\left\lbrace i^{(n)} ,j^{(n)} ,k^{(n)}\right\rbrace $ (see Eqs.~\eqref{eq:abcd-1},~\eqref{eq:abcd-2} and~\eqref{eq:abcd-3}), and similarly for $\left\lbrace a'^{(n)}_{1/2}, b'^{(n)}_{1/2}, c'^{(n)}_{1/2}, d'^{(n)}_{1/2}\right\rbrace $.
The summation over $s^{(n)}=1,2$ is simply introduced here to distinguish the two terms within the bracket.

In the second equality in Eq.~\eqref{eq:UUUU-1}, terms involving $\delta_{k^{(n)}_{1},k'^{(n)}_1}\delta_{k^{(n)}_{2},k'^{(n)}_2}\delta_{k^{(n)}_{1},k'^{(n)}_2}\delta_{k^{(n)}_{2},k'^{(n)}_1}$,
compared with those involving $\left( \delta_{k^{(n)}_{1},k'^{(n)}_1}\delta_{k^{(n)}_{2},k'^{(n)}_2}\right)^2 $ or 
$\left( \delta_{k^{(n)}_{1},k'^{(n)}_2}\delta_{k^{(n)}_{2},k'^{(n)}_1}\right)^2 $,
are of higher order in $1/q$, 
since the number of free summations is reduced.
For this reason, terms with $s^{(n)}\neq s^{(n+1)}$ for at least one $n$, compared with those with all $s^{(n)}$ identical, are of higher order in the large $q$ expansion and can be ignored.
As a result, we obtain
 \begin{align}\label{eq:UUUU-2}
\begin{aligned}
&\left\langle  
U_{\vex{i}_1 \vex{j}_1} U_{\vex{i}_2 \vex{j}_2} U^{\dagger}_{\vex{j}'_1 \vex{i}'_1} U^{\dagger}_{\vex{j}'_2 \vex{i}'_2} 
\right\rangle 
=
\sum_{\vex{k}_1,\vex{k}_1',\vex{k}_2,\vex{k}_2'}
\frac{1}{q^{4L'}}
\left(
\begin{aligned}
\prod_{n=1}^{L'}
&\delta_{a^{(n)}_1,a'^{(n)}_1}
\delta_{b^{(n+1)}_1,b'^{(n+1)}_1}
\delta_{c^{(n)}_1,c'^{(n)}_1}
\delta_{d^{(n+1)}_1,d'^{(n+1)}_1}
\\
&\times
\delta_{a^{(n)}_2,a'^{(n)}_2}
\delta_{b^{(n+1)}_2,b'^{(n+1)}_2}
\delta_{c^{(n)}_2,c'^{(n)}_2}
\delta_{d^{(n+1)}_2,d'^{(n+1)}_2}
\\
+
\prod_{n=1}^{L'}
&
\delta_{a^{(n)}_1,a'^{(n)}_2}
\delta_{b^{(n+1)}_1,b'^{(n+1)}_2}
\delta_{c^{(n)}_1,c'^{(n)}_2}
\delta_{d^{(n+1)}_1,d'^{(n+1)}_2}
\\
&\times
\delta_{a^{(n)}_2,a'^{(n)}_1}
\delta_{b^{(n+1)}_2,b'^{(n+1)}_1}
\delta_{c^{(n)}_2,c'^{(n)}_1}
\delta_{d^{(n+1)}_2,d'^{(n+1)}_1}
\end{aligned}
\right)
\\
&=
\frac{1}{q^{2L}}
\left(
\prod_{n=1}^{L}
\delta_{i_1^{(n)},i_1'^{(n)}}
\delta_{i_2^{(n)},i_2'^{(n)}}
\delta_{j_1^{(n)},j_1'^{(n)}}
\delta_{j_2^{(n)},j_2'^{(n)}}
+
\prod_{n=1}^{L}
\delta_{i_1^{(n)},i_2'^{(n)}}
\delta_{i_2^{(n)},i_1'^{(n)}}
\delta_{j_1^{(n)},j_2'^{(n)}}
\delta_{j_2^{(n)},j_1'^{(n)}}
\right).
\end{aligned}
\end{align}
In the second equality, we used the fact that Eq.~\eqref{eq:UU-1} reduces to Eq.~\eqref{eq:UU-2} (Eq.~\eqref{eq:UU-3})  upon substituting Eqs.~\eqref{eq:abcd-1} and~\eqref{eq:abcd-2}
(Eqs.~\eqref{eq:abcd-1},~\eqref{eq:abcd-2} and~\eqref{eq:abcd-3}).
This proves Eq.~\eqref{eq:cond} for any Floquet operator described by Eq.~\eqref{eq:U-S} with arbitrary permutation $\sigma \in S_{L'}$.

\section{Generalization to higher dimensional Floquet quantum circuits}\label{sec:AppFRC-HD}

In this appendix, we generalize the calculation in appendix~\ref{sec:AppFRC} to an arbitrary high dimension $D$. More specifically, we consider now a  $D$-dimensional Floquet random quantum circuit which consists of a $D$-dimensional cubic lattice of qudits.
During one period,
each qudit is coupled to all its neighboring qudits at different time substeps by different two-qudit unitary gates. All these unitary gates are drawn randomly  and independently from the CUE ensemble. 
We will prove that, for any Floquet quantum circuit of this type with an arbitrary ordering of the two-qudit unitary gates,  and for both periodic and open boundary conditions, the second and fourth order moments of the Floquet operator $U$  take the form Eq.~\eqref{eq:cond}, same as the CUE ensemble of dimension $N=q^{L^D}$, with $q$ being the Hilbert space dimension of each qudit and $L$ the number of lattice sites in each direction. For simplicity, we consider the large $q\rightarrow \infty$ limit for the calculation of the fourth order moment.

In the following, we employ the notation that each qudit is labeled by a $D$-dimensional vector $\vex{n}$ whose $\alpha$-th component $n_{\alpha}$ represents the coordinate in the lattice in dimension $\alpha=1,...,D$.
The neighbors of qudit $\vex{n}$ are denoted by $\vex{n}\pm \vex{e}_\alpha$, with $\vex{e}_{\alpha}$ being the unit vector in dimension $\alpha$.
$n_{\alpha}$ is defined modulo $L$ for periodic boundary condition.
The many-body state is indexed by $\vex{i}$ and the single-particle state at site $\vex{n}$ is labeled by $i^{(\vex{n})}$.

As in the 1D case, the second order moment $\left\langle  U_{\vex{i}\vex{j}} U^{\dagger}_{\vex{j}'\vex{i}'} \right\rangle $ can be expressed as 
\begin{align}\label{eq:UU-1-HD}
\begin{aligned}
\left\langle U_{\vex{i} \vex{j}}U^{\dagger}_{\vex{j}'\vex{i}'}\right\rangle 
=&
\sum_{\vex{k},\vex{k}'}
\prod_{\vex{n}}'
\prod_{\alpha=1}^{D}
\left\langle 
w^{(\vex{n},\vex{n}+\vex{e}_{\alpha})}_{a^{(\vex{n},\alpha)},b^{(\vex{n}+\vex{e}_{\alpha},\alpha)}; c^{(\vex{n},\alpha)} ,d^{(\vex{n}+\vex{e}_{\alpha},\alpha)} }
(w^{(\vex{n},\vex{n}+\vex{e}_{\alpha})})^{\dagger}_{c'^{(\vex{n},\alpha)},d'^{(\vex{n}+\vex{e}_{\alpha},\alpha)}; a'^{(\vex{n},\alpha)} ,b'^{(\vex{n}+\vex{e}_{\alpha},\alpha)} }
\right\rangle 
\\
=&
\sum_{\vex{k},\vex{k}'}
\prod_{\vex{n}}'
\prod_{\alpha=1}^{D}
\frac{1}{q^2}
\delta_{a^{(\vex{n},\alpha)},a'^{(\vex{n},\alpha)}}
\delta_{b^{(\vex{n}+\vex{e}_{\alpha},\alpha)},b'^{(\vex{n}+\vex{e}_{\alpha},\alpha)}}
\delta_{c^{(\vex{n},\alpha)},c'^{(\vex{n},\alpha)}}
\delta_{d^{(\vex{n}+\vex{e}_{\alpha},\alpha)},d'^{(\vex{n}+\vex{e}_{\alpha},\alpha)}}
.
\end{aligned}
\end{align}
Here $w^{(\vex{n},\vex{n}+\vex{e}_{\alpha})}$ represents the random CUE matrix that couples the qudits at sites $\vex{n}$ and $\vex{n}+\vex{e}_{\alpha}$.
$\prod_{\vex{n}}'$ stands for $\prod_{n_1=1}^{L'}...\prod_{n_D=1}^{L'}$, where $L'$ is defined as $L$ ($L-1$) for periodic (open) boundary condition.
We use $k^{(\vex{n},m)}$ to denote the single-particle state for qudit $\vex{n}$ at a intermediate substep, and it needs to be summed over ($\sum_{k^{(\vex{n},m)}=1}^{q}$).
$\sum_{\vex{k}}$ is short for the product 
$\prod_{\vex{n}} \prod_{m=1}^{M_{\vex{n}}}\sum_{k^{(\vex{n},m)}=1}^{q}$
where 
$m$ runs over $m=1,2,...,M_{\vex{n}}$ and $\vex{n}$ runs over all sites in the lattice.
$M_{\vex{n}}+1$ is the number of neighbors for site $\vex{n}$.
For periodic boundary condition $M_{\vex{n}}=2D-1$, while for open boundary condition \begin{align}\label{eq:Mn}
	M_{\vex{n}}=2D-D_{\vex{n}}-1,
\end{align}
where $D_{\vex{n}}$ denotes the number of components of vector $\vex{n}$ satisfying $n_{\alpha}=1$ or $L$.
$\sum_{\vex{k}'}$ is defined in the same manner.
The row and column indices $\left\lbrace a^{(\vex{n},\alpha)}, b^{(\vex{n},\alpha)} , c^{(\vex{n},\alpha)},d^{(\vex{n},\alpha)} \right\rbrace $ and $\left\lbrace a'^{(\vex{n},\alpha)}, b'^{(\vex{n},\alpha)} , c'^{(\vex{n},\alpha)},d'^{(\vex{n},\alpha)} \right\rbrace $
label the single-particle state at site $\vex{n}$ at a certain time substep and depend on the ordering of the quantum gates that couples sites $\vex{n}$ with all its neighbors.

Consider first the periodic boundary condition.
For arbitrary ordering of the two-qudit unitary gates, the $4D$ numbers in the set 
$S_{\vex{n}}^{(1)}=\left\lbrace a^{(\vex{n},\alpha)}, b^{(\vex{n},\alpha)} , c^{(\vex{n},\alpha)},d^{(\vex{n},\alpha)} | \alpha=1,2,...,D\right\rbrace $ are always related to those in the set
$S_{\vex{n}}^{(2)}=\left\lbrace i^{(\vex{n})}, j^{(\vex{n})} , k^{(\vex{n},m)}, k^{(\vex{n},m)}|m=1,2,...,2D-1\right\rbrace $ by a permutation.
Row and column indices for $w^{\dagger}$ are defined in an analogous way.  In particular,  the numbers in the set $S_{\vex{n}}'^{(1)}=\left\lbrace a'^{(\vex{n},\alpha)}, b'^{(\vex{n},\alpha)} , c'^{(\vex{n},\alpha)},d'^{(\vex{n},\alpha)} | \alpha=1,2,...,D\right\rbrace $ are related to those in the set
$S_{\vex{n}}'^{(2)}=\left\lbrace i'^{(\vex{n})}, j'^{(\vex{n})} , k'^{(\vex{n},m)}, k'^{(\vex{n},m)}| m=1,2,...,2D-1\right\rbrace $ by the same permutation that connects $S_{\vex{n}}^{(1)}$ to $S_{\vex{n}}^{(2)}$.
That means, for arbitrary ordering of the two-qudit gates, the following identity is always satisfied
 \begin{align}
 \prod_{\alpha=1}^{D}
 \delta_{a^{(\vex{n},\alpha)},a'^{(\vex{n},\alpha)}}
 \delta_{b^{(\vex{n},\alpha)},b'^{(\vex{n},\alpha)}}
 \delta_{c^{(\vex{n},\alpha)},c'^{(\vex{n},\alpha)}}
 \delta_{d^{(\vex{n},\alpha)},d'^{(\vex{n},\alpha)}}
 =
 \delta_{i^{(\vex{n})},i'^{(\vex{n})}}
 \delta_{j^{(\vex{n})},j'^{(\vex{n})}}
 \left( 
 \prod_{m=1}^{2D-1}\delta_{k^{(\vex{n},m)},k'^{(\vex{n},m)}}\right)^2.
 \end{align}
Inserting this equation into the second equality of Eq.~\eqref{eq:UU-1-HD}, we obtain the second order moment for the Floquet operator
\begin{align}\label{eq:UU-2-HD}
\begin{aligned}
\left\langle U_{\vex{i} \vex{j}}U^{\dagger}_{\vex{j}'\vex{i}'}\right\rangle 
=&
\sum_{\vex{k},\vex{k}'}
\prod_{\vex{n}}
\frac{1}{q^{2D}}
 \delta_{i^{(\vex{n})},i'^{(\vex{n})}}
\delta_{j^{(\vex{n})},j'^{(\vex{n})}}
\left(\prod_{m=1}^{2D-1}\delta_{k^{(\vex{n},m)},k'^{(\vex{n},m)}}\right)^2
%=
%\prod_{\vex{n}}'
%(\frac{q^{2D-1}}{q^{2D}})^{L^D}
%\delta_{i^{(\vex{n})},i'^{(\vex{n})}}
%\delta_{j^{(\vex{n})},j'^{(\vex{n})}}
\\
=&
\frac{1}{q^{L^D}}
\prod_{\vex{n}}
\delta_{i^{(\vex{n})},i'^{(\vex{n})}}
\delta_{j^{(\vex{n})},j'^{(\vex{n})}}
.
\end{aligned}
\end{align}

For the open boundary condition, extra care must be taken for the sites $\vex{n}$ at the boundaries.
For any site $\vex{n}$ sits in the bulk of the lattice,  as in the case of periodic boundary condition, the sets $S_{\vex{n}}^{(1)}$ and $S_{\vex{n}}^{(2)}$  ($S_{\vex{n}}'^{(1)}$ and $S_{\vex{n}}'^{(2)}$) defined above are always related by a $4D$-number permutation for any ordering of the quantum gates.
Moreover, the permutation that connects $S_{\vex{n}}^{(1)}$ and $S_{\vex{n}}^{(2)}$  is identical to the one that connects $S_{\vex{n}}'^{(1)}$ and $S_{\vex{n}}'^{(2)}$.
By contrast, at the boundary, for any component of the vector $\vex{n}$ that satisfies $n_{\alpha}=1$ ($n_{\alpha}=L$),   $b^{(\vex{n},\alpha)} $ and $d^{(\vex{n},\alpha)} $ ($a^{(\vex{n},\alpha)} $ and $c^{(\vex{n},\alpha)} $) have to be removed 
from the set
$S_{\vex{n}}^{(1)}=\left\lbrace a^{(\vex{n},\beta)}, b^{(\vex{n},\beta)} , c^{(\vex{n},\beta)},d^{(\vex{n},\beta)} | \beta=1,2,...,D\right\rbrace $.
After applying this procedure, all the $2M_{\vex{n}}+2$ numbers left in the new set $S_{\vex{n}}^{(1)}$ are related to those in the set $S_{\vex{n}}^{(2)}=\left\lbrace i^{(\vex{n})}, j^{(\vex{n})} , k^{(\vex{n},m)}, k^{(\vex{n},m)}|m=1,2,...,M_{\vex{n}}\right\rbrace $ by a permutation.
$\left\lbrace a'^{(\vex{n},\alpha)}, b'^{(\vex{n},\alpha)} , c'^{(\vex{n},\alpha)},d'^{(\vex{n},\alpha)} \right\rbrace $ are defined in the same way. They are related to  $\left\lbrace i'^{(\vex{n})}, j'^{(\vex{n})} , k'^{(\vex{n},m)}\right\rbrace $ in exactly  the same way
as $\left\lbrace a^{(\vex{n},\alpha)}, b^{(\vex{n},\alpha)} , c^{(\vex{n},\alpha)},d^{(\vex{n},\alpha)} \right\rbrace $ are related to  $\left\lbrace i^{(\vex{n})}, j^{(\vex{n})} , k^{(\vex{n},m)}\right\rbrace $.
This leads to 
 \begin{align}
 \begin{aligned}
& \prod_{\alpha | n_{\alpha }\neq L}
\delta_{a^{(\vex{n},\alpha)},a'^{(\vex{n},\alpha)}}
\delta_{c^{(\vex{n},\alpha)},c'^{(\vex{n},\alpha)}}
\prod_{\beta | n_{\beta }\neq 1}
\delta_{b^{(\vex{n},\beta)},b'^{(\vex{n},\beta)}}
\delta_{d^{(\vex{n},\beta)},d'^{(\vex{n},\beta)}}
\\
&=
\delta_{i^{(\vex{n})},i'^{(\vex{n})}}
\delta_{j^{(\vex{n})},j'^{(\vex{n})}}
\left( \prod_{m=1}^{M_{\vex{n}}}\delta_{k^{(\vex{n},m)},k'^{(\vex{n},m)}}\right)^2.
\end{aligned}
\end{align}
Here the first (second) product $\prod_{\alpha | n_{\alpha }\neq L}$ ($\prod_{\beta| n_{\beta }\neq 1}$) runs over all directions $\alpha$ ($\beta$) that obey $n_{\alpha}\neq L$ ($n_{\beta}\neq 1$).

Using this result, we find that Eq.~\eqref{eq:UU-1-HD} reduces to
\begin{align}\label{eq:UU-3-HD}
\begin{aligned}
\left\langle U_{\vex{i} \vex{j}} U^{\dagger}_{\vex{j}'\vex{i}'} \right\rangle 
=&
\sum_{\vex{k},\vex{k}'}
\prod_{\vex{n}}
\left( 
\frac{1}{q^{M_{\vex{n}}+1}}
\!\!\!\!
\prod_{\alpha | n_{\alpha }\neq L}
\!\!\!\!
\delta_{a^{(\vex{n},\alpha)},a'^{(\vex{n},\alpha)}}
\delta_{c^{(\vex{n},\alpha)},c'^{(\vex{n},\alpha)}}
\!\!\!\!
\prod_{\beta | n_{\beta }\neq 1}
\!\!\!\!
\delta_{b^{(\vex{n},\beta)},b'^{(\vex{n},\beta)}}
\delta_{d^{(\vex{n},\beta)},d'^{(\vex{n},\beta)}}
\right) 
\\
=&
\prod_{\vex{n}}
\left[ 
\frac{1}{q^{M_{\vex{n}}+1}}
\delta_{i^{(\vex{n})},i'^{(\vex{n})}}
\delta_{j^{(\vex{n})},j'^{(\vex{n})}}
\prod_{m=1}^{M_{\vex{n}}}
\left( 
\sum_{k^{(\vex{n},m)},k'^{(\vex{n},m)}}
\delta_{k^{(\vex{n},m)},k'^{(\vex{n},m)}}^2
\right)
\right] 
\\
=&
\frac{1}{q^{L^D}}
\prod_{\vex{n}}
\delta_{i^{(\vex{n})},i'^{(\vex{n})}}
\delta_{j^{(\vex{n})},j'^{(\vex{n})}}
.
\end{aligned}
\end{align}
%equivalent to the second order cumulant of the unitary matrix drawn from the CUE ensemble of dimension  $N=q^{L^D}$.
We have therefore proved Eq.~\eqref{eq:cond} for the $D$-dimensional Floquet random quantum circuit which has an arbitrary ordering of the two-qudit random unitary gates and is subject to either periodic or open boundary condition.

The fourth order moment $\left\langle  
U_{\vex{i}_1 \vex{j}_1} U_{\vex{i}_2 \vex{j}_2} U^{\dagger}_{\vex{j}'_1\vex{i}'_1} U^{\dagger}_{\vex{j}'_2\vex{i}'_2} 
\right\rangle $, in the limit of $q \rightarrow \infty$, can be expressed in a form similar to the 1D expression Eq.~\eqref{eq:UUUU-1}:
\begin{align}\label{eq:UUUU-1-HD}
\begin{aligned}
&\left\langle  
U_{\vex{i}_1 \vex{j}_1} U_{\vex{i}_2 \vex{j}_2} U^{\dagger}_{\vex{j}'_1\vex{i}'_1} U^{\dagger}_{\vex{j}'_2\vex{i}'_2} 
\right\rangle 
=
\sum_{\vex{k}_1,\vex{k}_1',\vex{k}_2,\vex{k}_2'}
\prod_{\vex{n}}'
\prod_{\alpha=1}^{D}
\left\langle 
w^{(\vex{n},\vex{n}+\vex{e}_{\alpha})}_{a^{(\vex{n},\alpha)}_1,b^{(\vex{n}+\vex{e}_{\alpha},\alpha)}_1; c^{(\vex{n},\alpha)}_1,d^{(\vex{n}+\vex{e}_{\alpha},\alpha)}_1 }
w^{(\vex{n},\vex{n}+\vex{e}_{\alpha})}_{a^{(\vex{n},\alpha)}_2,b^{(\vex{n}+\vex{e}_{\alpha},\alpha)}_2; c^{(\vex{n},\alpha)}_2 ,d^{(\vex{n}+\vex{e}_{\alpha},\alpha)}_2 }
\right. 
\\
&\times
\left. 
(w^{(\vex{n},\vex{n}+\vex{e}_{\alpha})})^{\dagger}_{c'^{(\vex{n})}_1,d'^{(\vex{n}+\vex{e}_{\alpha})}_1; a'^{(\vex{n})}_1 ,b'^{(\vex{n}+\vex{e}_{\alpha})}_1 }
(w^{(\vex{n},\vex{n}+\vex{e}_{\alpha})})^{\dagger}_{c'^{(\vex{n})}_2,d'^{(\vex{n}+\vex{e}_{\alpha})}_2; a'^{(\vex{n})}_2 ,b'^{(\vex{n}+\vex{e}_{\alpha})}_2 }
\right\rangle 
\\
=&
\sum_{\vex{k}_1,\vex{k}_1',\vex{k}_2,\vex{k}_2'}
\prod_{\vex{n}}'
\prod_{\alpha=1}^{D}
\frac{1}{q^4}
\sum_{s^{(\vex{n},\alpha)}=1}^{2}
\left(
\begin{aligned}
&
\delta_{s^{(\vex{n},\alpha)},1}
\delta_{a^{(\vex{n},\alpha)}_1,a'^{(\vex{n},,\alpha)}_1}
\delta_{b^{(\vex{n}+\vex{e}_{\alpha},\alpha)}_1,b'^{(\vex{n}+\vex{e}_{\alpha},\alpha)}_1}
\delta_{c^{(\vex{n},\alpha)}_1,c'^{(\vex{n},\alpha)}_1}
\delta_{d^{(\vex{n}+\vex{e}_{\alpha},\alpha)}_1,d'^{(\vex{n}+\vex{e}_{\alpha},\alpha)}_1}
\\
&\times
\delta_{a^{(\vex{n},\alpha)}_2,a'^{(\vex{n},\alpha)}_2}
\delta_{b^{(\vex{n}+\vex{e}_{\alpha},\alpha)}_2,b'^{(\vex{n}+\vex{e}_{\alpha},\alpha)}_2}
\delta_{c^{(\vex{n},\alpha)}_2,c'^{(\vex{n},\alpha)}_2}
\delta_{d^{(\vex{n}+\vex{e}_{\alpha},\alpha)}_2,d'^{(\vex{n}+\vex{e}_{\alpha},\alpha)}_2}
\\
+
&\delta_{s^{(\vex{n},\alpha)},2}
\delta_{a^{(\vex{n},\alpha)}_1,a'^{(\vex{n},\alpha)}_2}
\delta_{b^{(\vex{n}+\vex{e}_{\alpha},\alpha)}_1,b'^{(\vex{n}+\vex{e}_{\alpha},\alpha)}_2}
\delta_{c^{(\vex{n},\alpha)}_1,c'^{(\vex{n},\alpha)}_2}
\delta_{d^{(\vex{n}+\vex{e}_{\alpha},\alpha)}_1,d'^{(\vex{n}+\vex{e}_{\alpha},\alpha)}_2}
\\
&\times
\delta_{a^{(\vex{n},\alpha)}_2,a'^{(\vex{n},\alpha)}_1}
\delta_{b^{(\vex{n}+\vex{e}_{\alpha},\alpha)}_2,b'^{(\vex{n}+\vex{e}_{\alpha},\alpha)}_1}
\delta_{c^{(\vex{n},\alpha)}_2,c'^{(\vex{n},\alpha)}_1}
\delta_{d^{(\vex{n}+\vex{e}_{\alpha},\alpha)}_2,d'^{(\vex{n}+\vex{e}_{\alpha},\alpha)}_1}
\end{aligned}
\right) .
\end{aligned}
\end{align}
Here indices
$\left\lbrace a^{(\vex{n},\alpha)}_{1/2}, b^{(\vex{n},\alpha)}_{1/2}, c^{(\vex{n},\alpha)}_{1/2}, d^{(\vex{n})}_{1/2}\right\rbrace $ are related to $\left\lbrace i^{(\vex{n})}_{1/2} ,j^{(\vex{n})}_{1/2} ,k^{(\vex{n},m)}_{1/2}\right\rbrace $ in the same way as $\left\lbrace a^{(\vex{n},\alpha)}, b^{(\vex{n},\alpha)}, c^{(\vex{n},\alpha)}, d^{(\vex{n})} \right\rbrace $ are related to $\left\lbrace i^{(\vex{n})} ,j^{(\vex{n})} ,k^{(\vex{n},m)} \right\rbrace $ (similarly for the indices of $w^{\dagger}$).
As before, $s^{(\vex{n},\alpha)}$ is introduced to distinguish the two terms in the bracket.

Similar to the 1D case, in the second equality in the equation above, %Eq.~\eqref{eq:UUUU-1-HD},
terms that contain the factor $\delta_{k^{(\vex{n},m)}_{1},k'^{(\vex{n},m)}_1}\delta_{k^{(\vex{n},m)}_{2},k'^{(\vex{n},m)}_2}
\delta_{k^{(\vex{n},m)}_{1},k'^{(\vex{n},m)}_2}\delta_{k^{(\vex{n},m)}_{2},k'^{(\vex{n},m)}_1}$ are of higher order in the large $q$ expansion, compared with those with the factor
$\left( \delta_{k^{(\vex{n},m)}_{1},k'^{(\vex{n},m)}_1}\delta_{k^{(\vex{n},m)}_{2},k'^{(\vex{n},m)}_2}\right) ^2$ 
or
$\left( \delta_{k^{(\vex{n},m)}_{1},k'^{(\vex{n},m)}_2}\delta_{k^{(\vex{n},m)}_{2},k'^{(\vex{n},m)}_1}\right) ^2$
 due to the reduced number of free summations.
For this reason, we only need to keep terms for which all $s^{(\vex{n},\alpha)}$ are identical in the leading order of the large $q$ expansion.
This leads to
\begin{align}\label{eq:UUUU-2-HD}
\begin{aligned}
&\left\langle  
U_{\vex{i}_1 \vex{j}_1} U_{\vex{i}_2 \vex{j}_2} U^{\dagger}_{\vex{j}'_1\vex{i}'_1} U^{\dagger}_{\vex{j}'_2\vex{i}'_2} 
\right\rangle 
\\
=&
\sum_{\vex{k}_1,\vex{k}_1',\vex{k}_2,\vex{k}_2'}
\left(
\begin{aligned}
&
\prod_{\vex{n}}'
\prod_{\alpha=1}^{D}
\frac{1}{q^4}
\delta_{a^{(\vex{n},\alpha)}_1,a'^{(\vex{n},\alpha)}_1}
\delta_{b^{(\vex{n}+\vex{e}_{\alpha},\alpha)}_1,b'^{(\vex{n}+\vex{e}_{\alpha},\alpha)}_1}
\delta_{c^{(\vex{n},\alpha)}_1,c'^{(\vex{n},\alpha)}_1}
\delta_{d^{(\vex{n}+\vex{e}_{\alpha},\alpha)}_1,d'^{(\vex{n}+\vex{e}_{\alpha},\alpha)}_1}
\\
&\times
\delta_{a^{(\vex{n},\alpha)}_2,a'^{(\vex{n},\alpha)}_2}
\delta_{b^{(\vex{n}+\vex{e}_{\alpha},\alpha)}_2,b'^{(\vex{n}+\vex{e}_{\alpha},\alpha)}_2}
\delta_{c^{(\vex{n},\alpha)}_2,c'^{(\vex{n},\alpha)}_2}
\delta_{d^{(\vex{n}+\vex{e}_{\alpha},\alpha)}_2,d'^{(\vex{n}+\vex{e}_{\alpha},\alpha)}_2}
\\
&+
\prod_{\vex{n}}'
\prod_{\alpha=1}^{D}
\frac{1}{q^4}
\delta_{a^{(\vex{n},\alpha)}_1,a'^{(\vex{n},\alpha)}_2}
\delta_{b^{(\vex{n}+\vex{e}_{\alpha},\alpha)}_1,b'^{(\vex{n}+\vex{e}_{\alpha},\alpha)}_2}
\delta_{c^{(\vex{n},\alpha)}_1,c'^{(\vex{n},\alpha)}_2}
\delta_{d^{(\vex{n}+\vex{e}_{\alpha},\alpha)}_1,d'^{(\vex{n}+\vex{e}_{\alpha},\alpha)}_2}
\\
&\times
\delta_{a^{(\vex{n},\alpha)}_2,a'^{(\vex{n},\alpha)}_1}
\delta_{b^{(\vex{n}+\vex{e}_{\alpha},\alpha)}_2,b'^{(\vex{n}+\vex{e}_{\alpha},\alpha)}_1}
\delta_{c^{(\vex{n},\alpha)}_2,c'^{(\vex{n},\alpha)}_1}
\delta_{d^{(\vex{n}+\vex{e}_{\alpha},\alpha)}_2,d'^{(\vex{n}+\vex{e}_{\alpha},\alpha)}_1}
\end{aligned}
\right)
\\
&=
\frac{1}{q^{2L^D}}
\left(
\prod_{\vex{n}}
\delta_{i_1^{(\vex{n})},i_1'^{(\vex{n})}}
\delta_{i_2^{(\vex{n})},i_2'^{(\vex{n})}}
\delta_{j_1^{(\vex{n})},j_1'^{(\vex{n})}}
\delta_{j_2^{(\vex{n})},j_2'^{(\vex{n})}}
+
\prod_{\vex{n}}
\delta_{i_1^{(\vex{n})},i_2'^{(\vex{n})}}
\delta_{i_2^{(\vex{n})},i_1'^{(\vex{n})}}
\delta_{j_1^{(\vex{n})},j_2'^{(\vex{n})}}
\delta_{j_2^{(\vex{n})},j_1'^{(\vex{n})}}
\right) .
\end{aligned}
\end{align}
In the second equality here, we made use of the intermediate results in the derivation of the second order moment (i.e., the fact that  Eq.~\eqref{eq:UU-1-HD} is identical to Eq.~\eqref{eq:UU-2-HD} and Eq.~\eqref{eq:UU-3-HD} for periodic and open boundary condition, respectively).

\section{Effective field theory of a noninteracting Floquet model}~\label{sec:nonint}

In this appendix, we compare the effective field theory 
of the Floquet random quantum circuits studied earlier (or equivalently the CUE ensemble)
with that of a noninteracting Floquet model, and show that the quartic fluctuations become important in the noninteracting case. 
We consider now a time-periodic noninteracting system whose many-body Floquet operator is simply given by the tensor product of all single-particle Floquet operator $w^{(n)}$:
\begin{align}\label{eq:U-nonint}
\begin{aligned}
U=w^{(1)} \otimes w^{(2)} \otimes ... \otimes w^{(L)}.
\end{aligned}
\end{align}
All $\left\lbrace w^{(n)}| n=1,...,L\right\rbrace $ are $q \times q$ unitary matrices drawn randomly and independent from the CUE ensemble.
As before, we consider the limit of large single-particle Hilbert space dimension $q\rightarrow \infty$ for simplicity.

% \begin{figure}[tbp]
% 	\centering % \begin{center}/\end{center} takes some additional vertical space
% 	\includegraphics[width=.45\textwidth]{P2.png}
% 	\caption{\label{fig:ii} Floquet operator for a noninteracting Floquet quantum circuit. During one period, each qudit (indicated by a black dot) evolves independently and its time evolution within one period is generated by an independent random CUE matrix $w^{(n)}$  (represented by the blue box).   }
% \end{figure}

Since all single-particle Floquet operators $ w^{(n)}$ are independent, the many-body SFF $K(t)$ is given by the product of all single-particle SFFs $k^{(n)}(t)$:
\begin{align}\label{eq:nonint-SFF}
\begin{aligned}
K(t)
=
\prod_{n=1}^{L}k^{(n)}(t),
\qquad
k^{(n)}(t)=\left\langle |\Tr (w^{(n)})^t|^2\right\rangle .
\end{aligned}
\end{align}
Here $k^{(n)}(t)$ is simply the SFF for the CUE ensemble of dimension $q$ (Eq.~\eqref{eq:SFF-CUE} after the replacement $N\rightarrow q$). Therefore, one can easily see that the many-body SFF now exhibits a fast growing ramp $K(t)=t^L$ which plateaus at time $t=q$. 
The many-body quasi-energies are uncorrelated at the energy separation of the order of many-body level spacing $\Delta \phi \sim 1/N$, with $N=q^L$ being the dimension of many-body Hilbert space~\cite{PRL}.

The many-body Floquet operator $U$ given by Eq.~\eqref{eq:U-nonint} satisfies Eq.~\eqref{eq:cond-2} but not Eq.~\eqref{eq:cond-4}. Its fourth order moment in the large $q$ limit takes the form, 
\begin{align}\label{eq:cond-4-nonint}
\begin{aligned}
&\left\langle  
U_{\vex{i}_1 \vex{j}_1} 
U_{\vex{i}_2\vex{j}_2}
U^{\dagger}_{\vex{j}_1'\vex{i}_1'} 
U^{\dagger}_{\vex{j}_2'\vex{i}_2'} 
\right\rangle 
=
\prod_{n=1}^{L}
\left\langle  
w^{(n)}_{i^{(n)}_1,j^{(n)}_1} 
w^{(n)}_{i^{(n)}_2,j^{(n)}_2} 
(w^{(n)})^{\dagger}_{j'^{(n)}_1,i'^{(n)}_1 }
(w^{(n)})^{\dagger}_{j'^{(n)}_2,i'^{(n)}_2 }
\right\rangle 
\\
=&
\frac{1}{q^{2L}}
\prod_{n=1}^{L}
\left( 
\delta_{i^{(n)}_1,i'^{(n)}_1}
\delta_{j^{(n)}_1,j'^{(n)}_1}
\delta_{i^{(n)}_2,i'^{(n)}_2}
\delta_{j^{(n)}_2,j'^{(n)}_2}
+
\delta_{i^{(n)}_1,i'^{(n)}_2}
\delta_{j^{(n)}_1,j'^{(n)}_2}
\delta_{i^{(n)}_2,i'^{(n)}_1}
\delta_{j^{(n)}_2,j'^{(n)}_1}
\right) .
\end{aligned}
\end{align} 
Note that for the chaotic Floquet circuits described by Eq.~\eqref{eq:U-S}, to have nonzero moment
$\left\langle  
U_{\vex{i}_1 \vex{j}_1} 
U_{\vex{i}_2\vex{j}_2}
U^{\dagger}_{\vex{j}_1'\vex{i}_1'} 
U^{\dagger}_{\vex{j}_2'\vex{i}_2'} 
\right\rangle $,
if $i^{(n)}_1=i'^{(n)}_{1/2}$ is obeyed by one of the site, it must holds for all other sites as well. Otherwise, the fourth order moment can be ignored in the leading order of the large $q$ expansion.  
However, this is no longer the case for the noninteracting model. It is this difference that leads to the contrasting behaviors of the many-body level statistics for the two models.

Since the second order moment of the Floquet operator 
$\left\langle  
U_{\vex{i} \vex{j}} 
U^{\dagger}_{\vex{j}'\vex{i}'} 
\right\rangle$
for the current noninteracting model is identical to that of the chaotic Floquet circuits discussed in section~\ref{sec:FRC}, the quadratic fluctuations in the effective theory Eq.~\eqref{eq:Seff} are therefore also governed by the action in Eq.~\eqref{eq:Seff-2c}. 
On the other hand, the quartic order effective action of the noninteracting model is different from that of the chaotic model (Eq.~\eqref{eq:Seff-4c-XY}) and assumes the form
\begin{align}\label{eq:Seff4-nonint}
\begin{aligned}
S_{\rm eff}^{(4-1)}
=&
-
\frac{(\alpha\beta)^2}{2} 
\sum_{\ib_1,...,\ib_8}
\tr \left(Z^{\dagger}_{\ib_1 \ib_2} Z_{\ib_3 \ib_4}  \right)
\tr \left(Z^{\dagger}_{\ib_5 \ib_6}  Z_{\ib_7 \ib_8}  \right)
\\
&\times
\frac{1}{q^{2L}}
\left[
\begin{aligned}
&\prod_{n=1}^{L}
\left( 
\delta_{i^{(n)}_2,i^{(n)}_1}
\delta_{i^{(n)}_6,i^{(n)}_5}
\delta_{i^{(n)}_3,i^{(n)}_4}
\delta_{i^{(n)}_7,i^{(n)}_8}
+
\delta_{i^{(n)}_2,i^{(n)}_5}
\delta_{i^{(n)}_6,i^{(n)}_1}
\delta_{i^{(n)}_3,i^{(n)}_8}
\delta_{i^{(n)}_7,i^{(n)}_4}
\right) 
\\
&
-
\prod_{n=1}^{L}
\delta_{i^{(n)}_2,i^{(n)}_1}
\delta_{i^{(n)}_6,i^{(n)}_5}
\delta_{i^{(n)}_3,i^{(n)}_4}
\delta_{i^{(n)}_7,i^{(n)}_8}
\end{aligned}
\right] 
.
\end{aligned}
\end{align}
Here we have inserted Eq.~\eqref{eq:cond-4-nonint} into $S_{\rm eff}^{(4)}$ in Eq.~\eqref{eq:Seff4}, and, as before, ignored the last term in $S_{\rm eff}^{(4)}$ which is not important in the replica limit $R\rightarrow 0$.

Rewriting Eq.~\eqref{eq:Seff4-nonint} in terms of  $X$ and $Y$, which denote, respectively, the diagonal and off-diagonal components of the matrix $Z$ in the Hilbert space, we obtain
\begin{align}\label{eq:Seff-1-nonint}
\begin{aligned}
&S_{\rm eff}^{(4-1)}
=
-\frac{(\alpha\beta)^2}{2N^{2}} 
\sum_{\kb_1,...,\kb_4}
\tr \left(X^{\dagger}(\kb_1) X(\kb_2)  \right)
\tr \left(X^{\dagger}(\kb_3) X(\kb_4) \right)
\\
&\times
\left[
\prod_{n=1}^{L}
\left( 
\delta_{k^{(n)}_1,0}\delta_{k^{(n)}_2,0}\delta_{k^{(n)}_3,0}\delta_{k^{(n)}_4,0}
+
\frac{1}{q^2}\delta_{k^{(n)}_1,-k^{(n)}_3}\delta_{k^{(n)}_2,-k^{(n)}_4}
\right) 
-
\prod_{n=1}^{L}
\left( 
\delta_{k^{(n)}_1,0}\delta_{k^{(n)}_2,0}\delta_{k^{(n)}_3,0}\delta_{k^{(n)}_4,0}
\right) 
\right] 
\\
&-
\frac{(\alpha\beta)^2}{2N^2} 
\sum_{\ib_1\neq \ib_2,\ib_3\neq \ib_4,\ib_5\neq \ib_6, \ib_7 \neq \ib_8}
\tr \left(Y^{\dagger}_{\ib_1 \ib_2} Y_{\ib_3 \ib_4}  \right)
\tr \left(Y^{\dagger}_{\ib_5 \ib_6}  Y_{\ib_7 \ib_8}  \right)
\\
&
\times
\prod_{n=1}^{L}
\left( 
\delta_{i^{(n)}_2,i^{(n)}_1}
\delta_{i^{(n)}_6,i^{(n)}_5}
\delta_{i^{(n)}_3,i^{(n)}_4}
\delta_{i^{(n)}_7,i^{(n)}_8}
+
\delta_{i^{(n)}_2,i^{(n)}_5}
\delta_{i^{(n)}_6,i^{(n)}_1}
\delta_{i^{(n)}_3,i^{(n)}_8}
\delta_{i^{(n)}_7,i^{(n)}_4}
\right) 
\\
&-
\frac{(\alpha\beta)^2}{2N^{2}} 
\sum_{\ib_1\neq \ib_2,\ib_5\neq \ib_6}\sum_{\kb_3,\kb_4}
\left[ 
\tr \left(Y^{\dagger}_{\ib_1 \ib_2} X(\kb_3)  \right)
\tr \left(Y^{\dagger}_{\ib_5 \ib_6} X(\kb_4)  \right)
+
\tr \left(X^{\dagger}(-\kb_3) Y_{\ib_1 \ib_2}  \right)
\tr \left(X^{\dagger}(-\kb_4) Y_{\ib_5 \ib_6}  \right)
\right] 
\\
&\times
\prod_{n=1}^{L}
\left( 
\delta_{i^{(n)}_2,i^{(n)}_1}
\delta_{i^{(n)}_6,i^{(n)}_5}
\delta_{k^{(n)}_3,0} \delta_{k^{(n)}_4,0}
+
\frac{1}{q}
\delta_{i^{(n)}_2,i^{(n)}_5}
\delta_{i^{(n)}_6,i^{(n)}_1}
\delta_{k^{(n)}_3,-k^{(n)}_4}
\right) .
\end{aligned}
\end{align}
Here $X(\kb)$ represents the Fourier transform of the diagonal component $X_{\ib\ib}$ with respect to $\ib$ (Eq.~\eqref{eq:FT}), and the $n$-th component of momentum $\kb$ is denoted by $k^{(n)}$.

From Eq.~\eqref{eq:Seff-1-nonint}, we obtain the self-energy for $X(\kb)$ and $Y$ from the quartic interactions, in the replica limit $R\rightarrow 0$,
\begin{align}\label{eq:Sigma-nonint}
\begin{aligned}
(\Sigma_X)^{ab;ba}(\kb,\kb)
=&
\frac{(\alpha\beta)^2}{N^{2}} 
\left\langle 
X^{ab} (-\kb) 
(X^{\dagger})^{ba} (-\kb) 
\right\rangle_0 
\left[
\prod_{n=1}^{L}
\left( 
\delta_{k^{(n)},0}
+
\frac{1}{q^2}
\right) 
-
\delta_{\kb,0}
\right] 
\\
=&
\frac{(\alpha\beta)^2}{N} 
\frac{1}{1-\alpha\beta \delta_{\kb,0}}
\left[
\left( 
1
+
\frac{1}{q^2}
\right)^{r_0} 
\frac{1}{q^{2(L-r_0)}}
-
\delta_{\kb,0}
\right],
\\
(\Sigma_Y)^{ab;ba}_{\ib_1 \ib_2;\ib'_2 \ib'_1}
=&
\frac{(\alpha\beta)^2}{N^2} 
\sum_{\ib_3\neq \ib_4}
\left\langle 
Y^{ab}_{\ib_3 \ib_4}  
(Y^{\dagger})^{ba}_{\ib_4 \ib_3}  
\right\rangle_0 
\\
&
\times
\prod_{n=1}^{L}
\left( 
\delta_{i^{(n)}_2,i^{(n)}_1}
\delta_{i^{(n)}_3,i^{(n)}_4}
\delta_{i_1'^{(n)},i_2'^{(n)}}
+
\delta_{i^{(n)}_2,i^{(n)}_3}
\delta_{i^{(n)}_1,i^{(n)}_4}
\delta_{i_2'^{(n)},i^{(n)}_3}
\delta_{i_1'^{(n)},i^{(n)}_4}
\right) 
\\
%=&
%\frac{(\alpha\beta)^2}{N^2} 
%\left( \sum_{\ib_3, \ib_4}-\sum_{\ib_3=\ib_4}\right) 
%\prod_{n=1}^{L}
%\left( 
%\delta_{i^{(n)}_2,i^{(n)}_1}
%\delta_{i^{(n)}_3,i^{(n)}_4}
%\delta_{i_1'^{(n)},i_2'^{(n)}}
%+
%\delta_{i^{(n)}_2,i^{(n)}_3}
%\delta_{i^{(n)}_1,i^{(n)}_4}
%\delta_{i_2'^{(n)},i^{(n)}_3}
%\delta_{i_1'^{(n)},i^{(n)}_4}
%\right) 
%\\
=&
\frac{(\alpha\beta)^2}{N} 
\prod_{n=1}^{L}
\left( 
\delta_{i^{(n)}_2,i^{(n)}_1}
\delta_{i_1'^{(n)},i_2'^{(n)}}
+
\frac{1}{q}
\delta_{i^{(n)}_2,i_2'^{(n)}}
\delta_{i^{(n)}_1,i_1'^{(n)}}
\right) 
\\
&
-
\frac{(\alpha\beta)^2}{N} 
\prod_{n=1}^{L}
\left( 
\delta_{i^{(n)}_2,i^{(n)}_1}
\delta_{i_1'^{(n)},i_2'^{(n)}}
+
\frac{1}{q}
\delta_{i^{(n)}_2,i_2'^{(n)}}
\delta_{i^{(n)}_1,i_1'^{(n)}}
\delta_{i^{(n)}_1,i^{(n)}_2}
\right) .
\end{aligned}
\end{align}
Here we have used the expressions for the bare propagators Eqs.\eqref{eq:Prop-X} and~\eqref{eq:Prop-XY}. $r_0$ represents the number of zero components in the L-dimensional vector $\kb$. 
We can therefore see that the self-energy for $X(0)$ is of the order of  $L/q^2N(1-\exp(i\Delta\phi))$, and the self-energy for $X(\kb)$ with $r_0<L$ zero components in $\kb$ is of the order of  $1/q^{2(L-r_0)}N$.
Furthermore,  the self-energy $(\Sigma_Y)^{ab;ba}_{\ib_1\ib_2;\ib'_2\ib'_1}$ is of the order of $1/(Nq^{L-r'})$ where $r'$ represents the number of components satisfying $i_2^{(n)}=i_1^{(n)}$ and $i_2'^{(n)}=i_1'^{(n)}$.
Comparing Eqs.~\eqref{eq:Sigma-nonint} with  Eq.~\eqref{eq:Sigma-c}, we find the self-energies for $X(\kb)$ (with at least one zero component in $\kb$) and $Y_{\ib_1\ib_2}$ (when $i_1^{(n)}=i_2^{(n)}$ is satisfied for as least one site $n$)  of the current noninteracting model are much larger compared with their couterparts of the chaotic Floquet circuits.
%In particular,  for $\Delta \phi \lesssim \sqrt{L}/q$, the self-energy  is of the same order as the inverse of the bare propagator for $X(0)$.
%For the same reason that the higher order terms in the expansion of $R_2(\Delta \phi)$ in Eq.~\eqref{eq:nonint-R2-2} are nonnegligible, we can see that the quartic order corrections are important for the current noninteracting model.
Therefore, in contrast to the chaotic case, the quartic order fluctuations for the current noninteracting model are essential. 

\section{Large $q$ expansion of the Weingarten function}\label{sec:Wg-expan}

In this appendix, we study the asymptotic behavior of the Weingarten function $g(\sigma,\tau)$ (Eq.~\eqref{eq:g}) for the unitary group $U(q)$ in the large $q\rightarrow \infty$ limit.
Expanding the action $S_w[W,W^{\dagger}]$  in powers of $W$, we obtain
\begin{align}\label{eq:S-exp}
\begin{aligned}
& S_w[W,W^{\dagger}]
=
q
\sum_{n=1}^{\infty} \frac{(-1)^{n-1}}{n}
\tr \left( W W^{\dagger} \right)^{n},
\end{aligned}
\end{align}
where we have approximated the overall coefficient as $q+2R' \approx q$.
One can see from rescaling $W$ by $W \rightarrow W/\sqrt{q}$ that the higher order terms in the expansion above  are also of higher order in $1/q$.

To the leading order in the large $q$ expansion, we can therefore apply the Gaussian approximation and keep only the quadratic term in $S_w[W,W^{\dagger}]$: 
\begin{align}
S_w^{(2)}[W,W^{\dagger}] 
=
q \tr \left( W W^{\dagger} \right).
\end{align}
The bare $W$ propagator is given by 
\begin{align}
\begin{aligned}
\left\langle 
W_{uv}
W^{\dagger}_{v'u'}
\right\rangle_{0}
=\delta_{uu'}\delta_{vv'}
\frac{1}{q},
\end{aligned}
\end{align}
where the angular bracket with subscript $0$ denotes the functional averaging with the Gaussian action $S_w^{(2)}$.
Applying the Wick theorem to evaluate the expectation value of $\left\langle G[W,W^{\dagger} ;\sigma, \tau]\right\rangle_{0}  $ (Eq.~\eqref{eq:g}), we find that, in the Gaussian approximation, $g(\sigma,\tau)$ is nonvanishing only when $\sigma=\tau$:
\begin{align}
\begin{aligned}
&g(\sigma,\sigma)
=
\prod_{k=1}^{p}
\left\langle 
W_{k \sigma(k)}
W^{\dagger}_{\sigma(k) k}
\right\rangle_{0}
=
\frac{1}{q^p},
\end{aligned}
\end{align}
consistent with the known result for the Gaussian approximation to the Weingarten function $\mathrm{Wg}(id)=g(\sigma,\sigma)$~\cite{Beenakker}.
The Weingarten functions for all the remaining permutations $\tau^{-1}\sigma\neq id$ are of higher order in $1/q$.

To deduce the leading order term in the large $q$ expansion of $g(\sigma\neq \tau)$, we should go beyond the Gaussian approximation and keep the higher order terms in $S_w[W,W^{\dagger}]$.
Using Eq.~\eqref{eq:g-c}, we find 
\begin{align}
\begin{aligned}
& g(\sigma,\tau)
\propto 
\prod_{k=1}^{m}
\left\langle 
\left( 
\prod_{l=1}^{c_k}
W_{b_k+l,b_k+(l+1 \Mod c_k) }
W^{\dagger}_{b_k+(l+1 \Mod c_k), b_k+(l+1 \Mod c_k)}
\right) 
\left[ q \tr \left( W W^{\dagger} \right)^{c_k} \right]
\right\rangle_{0}
\\
&\propto 
\prod_{k=1}^{m}
\left( 
q
\prod_{l=1}^{c_k}
\left\langle 
W_{b_k+l,b_k+(l+1 \Mod c_k) }
W^{\dagger}_{b_k+(l+1 \Mod c_k),b_k+l}
\right\rangle_0
\right. 
\\
&\qquad \quad \times
\left. 
\left\langle  
W^{\dagger}_{b_k+(l+1 \Mod c_k), b_k+(l+1 \Mod c_k)}
W_{b_k+(l+1 \Mod c_k), b_k+(l+1 \Mod c_k)}
\right\rangle_0 
\right) 
\\
&\propto
\prod_{k=1}^{m}q^{1-2c_k},
\end{aligned}
\end{align}
which leads to the known result $g(\sigma,\tau)\propto q^{m-2p}$~\cite{Beenakker}.
The extra factor of $q \Tr \left( W W^{\dagger} \right)^{c_k}$ in the first line of the equation above represents the interaction vertex arising from the $c_k$-th order term in the expansion of $S_w[W,W^{\dagger}]$ (Eq.~\eqref{eq:S-exp}).

\section{Recursion relation for the Weingarten function}\label{sec:Wg-Rec}

In this appendix, we will show that the function $g(\sigma,\tau)$ given by Eq.~\eqref{eq:g} satisfies the recursion relations for the Weingarten function, i.e., Eq.~\eqref{eq:WgRec}.
To derive the first recursion relation Eq.~\eqref{eq:WgRec-a}, we make use of the  identity
$UU^{\dagger}=\bf{1}$,
which leads to
\begin{align}\label{eq:id-1}
\begin{aligned}
&
\int_{\mathcal{U}(q)} dU
U_{i_1 j_1}U_{i_2 j_2}...U_{i_{p} j_{p}}
U^{\dagger}_{j_1' i_1'}U^{\dagger}_{j_2' i_2'}...U^{\dagger}_{j'_{p} i'_{p}}
\\
=&
\!\!\!\!\!\!\!\!
\sum_{i_{p+1},i'_{p+1},j_{p+1}=1}^{q}
\!\!\!\!\!\!\!\!
\delta_{i_{p+1},i'_{p+1} }\delta_{j_{p+1},j'_{p+1}}
\int_{\mathcal{U}(q) } dU
U_{i_1 j_1} U_{i_2 j_2} ... U_{i_{p} j_{p}}U_{i_{p+1} j_{p+1}}
U^{\dagger}_{j_1' i_1'} U^{\dagger}_{j_2' i_2'}...U^{\dagger}_{j'_{p} i'_{p}}
U^{\dagger}_{j'_{p+1} i'_{p+1}}
.
\end{aligned}
\end{align}
Using Eq.~\eqref{eq:Wg0}, the left hand side (L.H.S.) of the equation above can be expressed as
\begin{align}\label{eq:LHS-1}
\begin{aligned}
&
\text{L.H.S}.
=
\sum_{\tau',\sigma' \in S_{p}}
\left( 
\prod_{k=1}^{p}
\delta_{i_k,i_{\sigma'(k)}'}\delta_{j_k,j_{\tau'(k)}'}
\right) 
g(\sigma',\tau'),
\end{aligned}
\end{align}
while the right hand side (R.H.S.) acquires the form
\begin{align}
\begin{aligned}
&
\text{R.H.S.}
=
\!\!\!\!\!\!\!\!
\sum_{i_{p+1},i'_{p+1},j_{p+1}=1}^{q}
\sum_{\tau'',\sigma'' \in S_{p+1}}
\delta_{i_{p+1},i'_{p+1} }\delta_{j_{p+1},j'_{p+1}}
\left( 
\prod_{k=1}^{p+1}
\delta_{i_k,i_{\sigma''(k)}'}\delta_{j_k,j_{\tau''(k)}'}
\right) 
g(\sigma'',\tau'')
.
\end{aligned}
\end{align}

We then choose an arbitrary pair of permutations $\sigma,\tau \in S_{p}$, and set $i_k=i_{\sigma (k)}', j_k=j_{\tau(k)}'$, for $k=1,2,...,p$. 
For simplicity, we consider the case where $p< q$ and  all integers in the set 
$\left\lbrace i'_k | k=1,...,p \right\rbrace $ ($\left\lbrace j'_k| k=1,...,p+1 \right\rbrace $) are different from each other.
In this case, the L.H.S. then becomes
\begin{align}\label{eq:LHS-2}
\begin{aligned}
&
\text{L.H.S.}
=
g(\sigma,\tau)
=
V^{(p)}_{c_1,c_2,...,c_m},
\end{aligned}
\end{align}
where $\left\lbrace c_k| k=1,2,...,m\right\rbrace $  represent the lengths of disjoint cycles for the permutation $\tau^{-1}\sigma$  and obey the constraint $\sum_{k=1}^{m}c_k=p$.
Here we made use of the result that $g(\sigma,\tau)$ depends only on $\left\lbrace c_k\right\rbrace $ and expressed it as
$V^{(p)}_{c_1,c_2,...,c_m}$ (see Sec.~\ref{sec:Wg-2}).
On the other hand, the R.H.S. becomes
\begin{align}\label{eq:RHS-1}
\begin{aligned}
\text{R.H.S.}
=&
\sum_{\tau'',\sigma'' \in S_{p+1}}
\sum_{i'_{p+1}=1}^{q}
\left( 
\prod_{k=1}^{p}
\delta_{i_{\sigma(k)}',i_{\sigma''(k)}'}
\delta_{j_{\tau(k)}',j_{\tau''(k)}'}
\right) 
%\\
%&\times
\delta_{i'_{p+1} ,i_{\sigma''(p+1)}'}\delta_{j'_{p+1},j_{\tau''(p+1)}'}
g(\sigma'',\tau'').
\end{aligned}
\end{align}

Let us now examine all terms in the summation in Eq.~\eqref{eq:RHS-1}.
The contribution is nonzero for arbitrary $i_{p+1}'$ if $\sigma''=\sigma_p$ and $\tau''=\tau_p$,  where
$\sigma_p$ and $\tau_p$ are defined as
\begin{align}\label{eq:taup}
\begin{aligned}
\sigma_p(k)
=
\begin{cases}
\sigma(k),
&
k=1,2,...,p,
\\
p+1,
&
k=p+1,
\end{cases}
\qquad
\tau_p(k)
=
\begin{cases}
\tau(k),
&
k=1,2,...,p,
\\
p+1,
&
k=p+1.
\end{cases}
\end{aligned}
\end{align}
The permutation $\tau^{-1}_p\sigma_p$, compared with $\tau^{-1}\sigma$, contains an extra disjoint cycle of length $1$.
More specifically, from earlier result (Eq.~\eqref{eq:g1}),  we find that $g(\sigma_p,\tau_p)$ is given by
\begin{align}
\begin{aligned}
&
g(\sigma_p,\tau_p)
=
(-1)^{p-m}
\left\langle 
\left( 
W_{p+1,p+1}
W^{\dagger}_{p+1,p+1}
\right) 
\prod_{k=1}^{m}
\left( 
\prod_{l=1}^{c_k}
W_{P_{l}^{(k)} , \sigma(P_{l}^{(k)})}
W^{\dagger}_{\sigma(P_{l}^{(k)}) , P_{l+1}^{(k)}}
\right) 
\right\rangle_{W} 
\\
=&
V^{(p+1)}_{c_1,c_2,...c_m,1}.
\end{aligned}
\end{align}
Compared with $g(\sigma,\tau)=V^{(p)}_{c_1,c_2,...c_m}$, it contains an extra factor of 
$
W_{p+1,p+1}
W^{\dagger}_{p+1,p+1}
$.
Here the integers $\left\lbrace  P_{l}^{(k)}\right\rbrace $ denote the cycle structure of  $\tau^{-1}\sigma$ as in Eq.~\eqref{eq:cycle}.
The total contribution from this pair of permutations ($\sigma_p,\tau_p$) to Eq.~\eqref{eq:RHS-1} is
\begin{align}\label{eq:WgRec-a-1}
\begin{aligned}
\sum_{i_{p+1}'=1}^{q} g(\sigma_p,\tau_p)
=
qV^{(p+1)}_{c_1,c_2,...,c_m,1},
\end{aligned}
\end{align}
where the factor $q$ comes from the summation over $i_{p+1}'$.

Another nonvanishing contribution comes from the case when $i_{p+1}'=i_{\sigma(l)}'$ for an arbitrary positive integer $l \leq p$. In this case, the contribution is nonvanishing if the permutations $\sigma''=\sigma_l$ and $\tau''=\tau_p$, where $\tau_p$ is defined in Eq.~\eqref{eq:taup} and $\sigma_l$ is given by
\begin{align}
\begin{aligned}
&
\sigma_l(k)
=
\begin{cases}
\sigma(k),
&
k \neq l ,p+1,
\\
p+1,
&
k = l,
\\
\sigma(l),
&
k=p+1.
\end{cases}
\end{aligned}
\end{align}
In this case, it is easy to see that $\delta_{i'_{p+1} ,i_{\sigma_l(p+1)}'}=\delta_{i'_{\sigma(l)},i_{\sigma_l(l)}'}=\delta_{i'_{p+1} ,i_{\sigma(l)}'}=1$.

$g( \sigma_l,\tau_p)$ can be obtained from  $g(\sigma,\tau)$ by replacing  the factor $W_{l,\sigma(l)} W^{\dagger}_{\sigma(l),\tau^{-1}\sigma(l)}$ in Eq.~\eqref{eq:g1}
with $	W_{l,p+1} 
W^{\dagger}_{p+1,p+1} 
W_{p+1,\sigma(l)}
W^{\dagger}_{\sigma(l),\tau^{-1}\sigma(l)}$.
This results in adding a number to the disjoint cycle which number $l$ belongs (labeled by $j$):
\begin{align}
\begin{aligned}
g(\sigma,\tau)
=&
\left\langle 
W_{l,\sigma(l)} W^{\dagger}_{\sigma(l),\tau^{-1}\sigma(l)}...
\right\rangle_W 
=
V_{c_1,...,c_j,...,c_m}^{(p)},
\\
g( \sigma_l,\tau_p)
=&
\left\langle 
W_{l,\sigma_l(l)} W^{\dagger}_{\sigma_l(l),\tau_p^{-1}\sigma_l(l)} 
W_{\tau_p^{-1}\sigma_l(l),\sigma_l\tau_p^{-1}\sigma_l(l)}
W^{\dagger}_{\sigma_l\tau_p^{-1}\sigma_l(l),\tau_p^{-1}\sigma_l\tau_p^{-1}\sigma_l(l)} ...
\right\rangle_W ,
\\
=&
\left\langle 
W_{l,p+1} 
W^{\dagger}_{p+1,p+1} 
W_{p+1,\sigma(l)}
W^{\dagger}_{\sigma(l),\tau^{-1}\sigma(l)} ...
\right\rangle_W 
=
V_{c_1,...,c_j+1,...,c_m}^{(p+1)}.
\end{aligned}
\end{align}
Here ``$...$'' represents all remaining terms that stay the same in $g( \sigma_l,\tau_p)$ and $g(\sigma,\tau)$.
Summing over all possibilities of $l$, we find the total contribution of this type is 
\begin{align}\label{eq:WgRec-a-2}
\begin{aligned}
\sum_{l} g(\sigma_l,\tau_p)
=
\sum_{s=1}^{m} c_s
V_{c_1,...,c_{s-1},c_{s}+1,c_{s+1},...,c_m}^{(p+1)}.
\end{aligned}
\end{align}

%Besides the abovementioned cases, all remaining terms in Eq.~\eqref{eq:RHS-1}  vanish.
Combining everything (Eqs.~\eqref{eq:WgRec-a-1} and~\eqref{eq:WgRec-a-2}), we find the R.H.S. of Eq.~\eqref{eq:id-1}
\begin{align}
\begin{aligned}
\text{R.H.S.}
=
qV_{c_1,...,c_m,1}^{(p+1)}
+
\sum_{s=1}^{m} c_s
V_{c_1,...,c_{s-1},c_s+1, c_{s+1},...,c_m}^{(p+1)}.
\end{aligned}
\end{align}
Equating this to the L.H.S. (Eq,~\eqref{eq:LHS-2}), we arrive at the recursion relation Eq.~\eqref{eq:WgRec-a} which holds for arbitrary pair of $\sigma,\tau \in S_{p}$ (or equivalently arbitrary configuration of $\left\lbrace c_j\right\rbrace $ that satisfies $\sum_{j=1}^{m}c_j=p$) when $p<q$.

%%%%%%%%%%%%%%%%%%%%%

The recursion relation Eq.~\eqref{eq:WgRec-b} can be proved in a similar way using the identity:
\begin{align}\label{eq:id-2}
\begin{aligned}
\sum_{j_p,j_p'=1}^{q}
\delta_{j_p,j_p'}
\int_{\mathcal{U}(q)} dU
U_{i_1 j_1}U_{i_2 j_2}
...
U_{i_{p-1} j_{p-1} }
U_{i_p j_p}
U^{\dagger}_{j_1' i_1'}U^{\dagger}_{j_2' i_2'}
...
U^{\dagger}_{j_{p-1}' i_{p-1}'}
U^{\dagger}_{j_p'  i_p'}
=
0,
\end{aligned}
\end{align}
for $i_p \neq i_p'$.
Using Eq.~\eqref{eq:Wg0}, one can rewrite this equation as
\begin{align}\label{eq:LHS-b}
\begin{aligned}
&
\sum_{j_p,j_p'=1}^{q}
\sum_{\tau',\sigma' \in S_p}
\delta_{j_p,j_p'}
\left( 
\prod_{k=1}^{p}
\delta_{i_k,i_{\sigma'(k)}'}\delta_{j_k,j_{\tau'(k)}'}
\right) 
g(\sigma',\tau')
=
0.
\end{aligned}
\end{align}

We now set $j_{k}=j_{k}'$  and $i_{k}=i_{\chi(k)}'$ for $1 \leq k \leq  p$. $\chi \in S_p$ represents an arbitrary nonidentical permutation of numbers $1,2,...,p$ which satisfies $\chi(p) \neq p$.
It contains $m$ disjoint cycles with lengths $\left\lbrace c_k| k=1,2,...,m\right\rbrace $. Since $\chi \neq id$ and $\chi(p) \neq p$, 
$p$ stays in a cycle (labeled by $1$) of length $c_1\geq 2$.
We consider the case $p \leq q$ and choose the set 
$\left\lbrace i'_k| k=1,...,p \right\rbrace $ ($\left\lbrace j'_k| k=1,...,p-1 \right\rbrace $) whose elements are all different from each other for simplicity.
Eq.~\eqref{eq:LHS-b} can be rewritten as
\begin{align}\label{eq:LHS-c}
\begin{aligned}
&
\sum_{j_p'=1}^{q}
\sum_{\tau',\sigma' \in S_p}
\left( 
\prod_{k=1}^{p}
\delta_{i_{\chi(k)}',i_{\sigma'(k)}'}
\delta_{j_k',j_{\tau'(k)}'}
\right) 
g(\sigma',\tau')
=
0.
\end{aligned}
\end{align}

Let us now examine all terms in the summation on the L.H.S. of Eq.~\eqref{eq:LHS-c}.
For arbitrary $j_p'$, the contribution is nonvanishing when $\sigma'=\chi$ and  $\tau'=id$. The total contribution of this type  is
\begin{align}\label{eq:WgRec-b-1}
\begin{aligned}
\sum_{j_p'=1}^{q}
 g(\chi, id)
=
qg(\chi, id)
=
qV_{c_1,c_2,...,c_m}^{(p)}.
\end{aligned}
\end{align}

Another type of nonvanishing contributions comes from the case where $j_p'=j_{l}'$ and $l$ also stays within the same cycle that $p$ belongs (cycle $1$). In this case, the contribution is nonvanishing if $\sigma'=\chi$ and $\tau=\tau_l$, where $\tau_{l}$ is defined as
\begin{align}\label{eq:taul}
\begin{aligned}
\tau_l(k)
=
\begin{cases}
k,
&
k\neq p ,l,
\\
l,
&
k=p,
\\
p,
&
k=l.
\end{cases}
\end{aligned}
\end{align}
$g(\chi,\tau_l)$ can be obtained from $g(\chi,id)$ by replacing $W^{\dagger}_{ll}$ and $W^{\dagger}_{pp}$ in $G[W,W^{\dagger}; \chi, id]$ (Eq.~\eqref{eq:g}) with $W^{\dagger}_{lp}$ and $W^{\dagger}_{pl}$.
This breaks the  cycle $1$ of length $c_1$ into two disjoint cycles of lengths $c_1-c$ and $c$ ($c$ obeys $0<c<c_1$ and depends on $l$):
\begin{align}
\begin{aligned}
& 
g(\chi,id)
=V^{(p)}_{c_1,c_2,...,c_m}
\\
=&
\left\langle 
\left(
W_{p, \chi(p)} W^{\dagger}_{\chi(p),\chi(p)}
...
W_{\chi^{-1}(l),l} W^{\dagger}_{l,l} 
W_{l,\chi(l)}W^{\dagger}_{\chi(l),\chi(l)}
...
W_{\chi^{-1}(p), p} W^{\dagger}_{p,p}
\right) 
\prod_{k\neq 1}^{m} (...)
\right\rangle_W,
\\
&
g(\chi,\tau_l)
=V^{(p)}_{c_1-c,c,c_2,...,c_m}
\\
=&
\left\langle 
\left(
W_{p, \chi(p)} W^{\dagger}_{\chi(p),\chi(p)}
...
W_{\chi^{-1}(l),l} W^{\dagger}_{l,p}
\right) 
\left( 
W_{l,\chi(l)}W^{\dagger}_{\chi(l),\chi(l)}
...
W_{\chi^{-1}(p), p} W^{\dagger}_{p,l}
\right) 
\prod_{k\neq 1}^{m} (...)
\right\rangle_W 
.
\end{aligned}
\end{align}
Here $\prod_{k\neq 1}^{m} (...)$ represents the contribution from all remaining cycles other than $1$, and stays the same for both $g(\chi,id)$ and $g(\chi,\tau_l)$.
Taking into account all possible $l$ belonging to the cycle $1$, we find that the total contribution from such terms is 
\begin{align}\label{eq:WgRec-b-2}
\begin{aligned}
\sum_{l \in \text{cycle 1} } g(\chi,\tau_l)
=
\sum_{c=1}^{c_1-1}
V_{c_1-c,c,c_2,...,c_m}^{(p)}.
\end{aligned}
\end{align}

The last type of nonvanishing contribution arises from the case when $j_p'=j_{l}'$ and $l$ does not belong to the cycle $1$ (we label the cycle which $l$ belongs by $2$). In this case, the contribution is nonvanishing if $\sigma'=\chi$ and $\tau'=\tau_l$ (Eq.~\eqref{eq:taul}).
As mentioned earlier,  $g(\chi,\tau_l)$ can be obtained from $g(\chi,id)$ by replacing $W^{\dagger}_{ll}$ and $W^{\dagger}_{pp}$ in $G[W,W^{\dagger}; \chi, id]$ with $W^{\dagger}_{lp}$ and $W^{\dagger}_{pl}$, respectively.
However, since now $l,p$ belong to different cycles, this replacement  combines the cycle $1$ and cycle $2$ into one cycle of length $c_1+c_2$:
\begin{align}
\begin{aligned}
&g(\chi,id)	=V_{c_1,c_2,c_3,...,c_m}^{(p)}
\\
=&
\left\langle 
\left(
W_{p, \chi(p)} W^{\dagger}_{\chi(p),\chi(p)}
...
W_{\chi^{-1}(p), p} W^{\dagger}_{p,p}
\right) 
%\right. 
%\\
%& \times 
%\left. 
\left( 
W_{l,\chi(l)}W^{\dagger}_{\chi(l),\chi(l)}
...
%W_{\chi^{c_2-2}(l),\chi^{-1}(l)} W^{\dagger}_{\chi^{-1}(l),\chi^{-1}(l)} 
W_{\chi^{-1}(l),l} W^{\dagger}_{l,l} 
\right) 
\prod_{k> 2}^{m} (...)
\right\rangle ,
\\
&g(\chi,\tau_0)	
=
V_{c_1+c_2,c_3,...,c_m}^{(p)}
\\
%=
%&\left\langle 
%\left(
%W_{p, \chi(p)} W^{\dagger}_{\chi(p)\chi(p)}
%...
%...
%W_{\chi^{c_1}(p), p} W^{\dagger}_{p,l}
%\right) 
%\left( 
%W_{\chi^{-1}(l),l} W^{\dagger}_{l,p} 
%W_{l,\chi(l)}W^{\dagger}_{\chi(l),\chi(l)}
%...
%W_{\chi^{c_2-1}(l),\chi^{-1}(l)} W^{\dagger}_{\chi^{-1}(l),\chi^{-1}(l)} 
%\right) 
%\prod_{j> 2} (...)
%\right\rangle 
%\\
=&
\left\langle 
\left(
W_{p, \chi(p)} W^{\dagger}_{\chi(p),\chi(p)}
...
W_{\chi^{-1}(p), p} W^{\dagger}_{p,l}
%\right. \right. 
%\\
%&\times
%\left. \left. 
W_{l,\chi(l)}W^{\dagger}_{\chi(l),\chi(l)}
...
%W_{\chi^{c_2-1}(l),\chi^{-1}(l)} W^{\dagger}_{\chi^{-1}(l),\chi^{-1}(l)} 
W_{\chi^{-1}(l),l} W^{\dagger}_{l,p} 
\right) 
\prod_{k> 2}^m (...)
\right\rangle .
\end{aligned}
\end{align}
Here $\prod_{k>2}^m (...)$ represents the contribution from all remaining cycles $k\neq 1,2$, which remains the same for both cases.
Summing over all possible $l$ that do not stay in cycle $1$, we find the total contribution of this type is
\begin{align}\label{eq:WgRec-b-3}
\begin{aligned}
\sum_{l \notin \text{cycle 1} } g(\chi,\tau_l)
=
\sum_{s=2}^{m}
c_s
V_{c_1+c_s,c_2,...,c_{s-1},c_{s+1},...,c_m}^{(p)}
\end{aligned}
\end{align}.

From Eq.~\eqref{eq:id-2}, one can see that, 
for arbitrary choice of $\chi\in S_p$ that satisfies $\chi(p)\neq p$ (for arbitrary $\left\lbrace c_k\right\rbrace $ with at least one cycle length $c_1>1$),
the sum of  all three types of nonvanishing contributions (Eqs.~\eqref{eq:WgRec-b-1}, \eqref{eq:WgRec-b-2} and~\eqref{eq:WgRec-b-3}) should equal $0$, which proves Eq.~\eqref{eq:WgRec-b} for $p\leq q$.


%\begin{thebibliography}{99}
%
%\bibitem{a}
%Author, \emph{Title}, \emph{J. Abbrev.} {\bf vol} (year) pg.
%
%\bibitem{b}
%Author, \emph{Title},
%arxiv:1234.5678.
%
%\bibitem{c}
%Author, \emph{Title},
%Publisher (year).
%
%
%% Please avoid comments such as "For a review'', "For some examples",
%% "and references therein" or move them in the text. In general,
%% please leave only references in the bibliography and move all
%% accessory text in footnotes.
%
%% Also, please have only one work for each \bibitem.
%
%
%\end{thebibliography}

\bibliographystyle{JHEP}
\bibliography{circuit}

\end{document}